\documentclass[traditabstract]{aa}
\usepackage{txfonts}
\usepackage{natbib}
\usepackage{longtable}
\usepackage{multirow}
\usepackage{color}
\usepackage{graphicx}

\newcommand{\sfrsed}{\ifmmode  \mathrm{SFR}_\mathrm{SED} \else $\mathrm{SFR}_\mathrm{SED}$\fi}
\newcommand{\sfruvir}{\ifmmode  \mathrm{SFR}_\mathrm{UV+IR} \else $\mathrm{SFR}_\mathrm{UV+IR}$\fi}
\newcommand{\sfruv}{\ifmmode  \mathrm{SFR}_\mathrm{UV} \else $\mathrm{SFR}_\mathrm{UV}$\fi}
\newcommand{\sfrlya}{\ifmmode  \mathrm{SFR}_{\mathrm{Ly}\alpha} \else $\mathrm{SFR}_{\mathrm{Ly}\alpha}$\fi}
\newcommand{\muv}{\ifmmode  \mathrm{M}_\mathrm{UV} \else $\mathrm{M}_\mathrm{UV}$\fi}
\newcommand{\sfrha}{\ifmmode  \mathrm{SFR}_{\mathrm{H}\alpha} \else $\mathrm{SFR}_{\mathrm{H}\alpha}$\fi}
\newcommand{\flyc}{\ifmmode  \mathrm{f}_\mathrm{esc}\mathrm{(LyC)} \else $\mathrm{f}_\mathrm{esc}\mathrm{(LyC)}$\fi}
\newcommand{\flya}{\ifmmode  \mathrm{f}_\mathrm{esc}\mathrm{(Ly\alpha)} \else $\mathrm{f}_\mathrm{esc}\mathrm{(Ly}\alpha)$\fi}
\newcommand{\fefflya}{\ifmmode  \mathrm{f}_\mathrm{esc}^\mathrm{eff}\mathrm{(Ly\alpha)} \else $\mathrm{f}_\mathrm{esc}^\mathrm{eff}\mathrm{(Ly}\alpha)$\fi}
\newcommand{\flyarel}{\ifmmode  \mathrm{f}_\mathrm{esc}^\mathrm{rel}\mathrm{(Ly\alpha)} \else $\mathrm{f}_\mathrm{esc}^\mathrm{rel}\mathrm{(Ly}\alpha)$\fi}
\newcommand{\dfel}{\ifmmode \Delta\log {\rm f}_{\rm EL} \else $\Delta\log$ f$_{\rm EL}$\fi}
\newcommand{\hst}{{\it HST}}
\newcommand{\jwst}{{\it JWST}}

\newcommand{\hyperz}{{\it Hyperz}}
\newcommand{\flyased}{\ifmmode  \mathrm{f}_\mathrm{SED}\mathrm{(Ly\alpha)} \else $\mathrm{f}_\mathrm{SED}\mathrm{(Ly}\alpha)$\fi}
\newcommand{\flyaobs}{\ifmmode  \mathrm{f}_\mathrm{obs}\mathrm{(Ly\alpha)} \else $\mathrm{f}_\mathrm{obs}\mathrm{(Ly}\alpha)$\fi}
\newcommand{\tigm}{\ifmmode  \mathrm{T}_\mathrm{IGM} \else $\mathrm{T}_\mathrm{IGM}$\fi}

\newcommand{\hi}{\textrm{H}~\textsc{i}}
\newcommand{\oiii}{[\textrm{O}~\textsc{iii}]}
\newcommand{\oii}{[\textrm{O}~\textsc{ii}]}
\newcommand{\oiilam}{[\textrm{O}~\textsc{ii}]\ensuremath{\lambda3727}}

\newcommand{\oiiidoub}{[\textrm{O}~\textsc{iii}]\ensuremath{\lambda\lambda4959,5007}}
\newcommand{\niilam}{[\textrm{N}~\textsc{ii}]\ensuremath{\lambda6583}} 
\newcommand{\ha}{\ifmmode {\rm H}\alpha \else H$\alpha$\fi}
\newcommand{\hb}{\ifmmode {\rm H}\beta \else H$\beta$\fi}
\newcommand{\lya}{\ifmmode {\rm Ly}\alpha \else Ly$\alpha$\fi}
\newcommand{\pg}{\ifmmode {\rm P}\gamma \else P$\gamma$\fi}
\newcommand{\lyb}{\ifmmode {\rm Ly}\beta \else Ly$\beta$\fi}
\newcommand{\lyg}{\ifmmode {\rm Ly}\gamma \else Ly$\gamma$\fi}
\newcommand{\ciii}{\textrm{C}~\textsc{iii}]\ensuremath{\lambda1909}}

\newcommand{\civ}{\textrm{C}~\textsc{iv}\ensuremath{\lambda1550}}
\newcommand{\heii}{\textrm{He}~\textsc{ii}\ensuremath{\lambda1640}}

\def\ergs{\ifmmode \mathrm{erg\hspace{1mm}s}^{-1} \else erg s$^{-1}$\fi}

\def\micron{\ifmmode \mu\mathrm{m} \else $\mu$m\fi}
\def\msun{\ifmmode \mathrm{M}_{\odot} \else M$_{\odot}$\fi}
\def\msunyr{\ifmmode \mathrm{M}_{\odot} \hspace{1mm}{\rm yr}^{-1} \else $\mathrm{M}_{\odot}$ yr$^{-1}$\fi}
\def\zsun{\ifmmode Z_{\odot} \else Z$_{\odot}$\fi}
\def\lsun{\ifmmode L_{\odot} \else L$_{\odot}$\fi}
\def\mstar{\ifmmode \mathrm{M}_{\star} \else M$_{\star}$\fi}

\newcommand{\myemail}{stephane.debarros@unige.ch}
\newcommand{\myinstitute}{Observatoire de Gen\`{e}ve, Universit\'{e} de Gen\`{e}ve, 51 Ch. des Maillettes, 1290 Versoix, Switzerland}

\begin{document}
\title{A VLT/FORS2 view at $z\sim6$: Lyman-$\alpha$ emitter fraction and galaxy physical properties at the edge of the epoch of cosmic reionization}

   \author{S.~De~Barros
   \inst{1,2}
   \and
   L.~Pentericci
   \inst{3}
   \and
   E.~Vanzella
   \inst{2}
   \and
   M.~Castellano
   \inst{3}
   \and
   A.~Fontana
   \inst{3}
   \and
   A.~Grazian
   \inst{3}
   \and
   C.~J.~Conselice
   \inst{4}
   \and
   H.~Yan
   \inst{5}
   \and
    A.~Koekemoer
   \inst{6}
   \and
   S.~Cristiani
   \inst{7}
   M.~Dickinson
   \inst{8}
   \and
   S.~L.~Finkelstein
   \inst{9}
   \and
   R.~Maiolino
   \inst{10,11}
   }

\institute{\myinstitute\\ \email{\myemail}
\and
INAF--Osservatorio Astronomico di Bologna, via Ranzani 1, 40127 Bologna, Italy
\and
INAF--Osservatorio Astronomico di Roma, via Frascati 33, 00040 Monteporzio, Italy
\and
School of Physics \& Astronomy, The University of Nottingham, University Park, Nottingham NG7 2RD, UK
\and
Department  of  Physics  and  Astronomy  University  of  Missouri Columbia, MO 65211, USA
\and
Space Telescope Science Institute, 3700 San Martin Drive, Baltimore, MD 21218, USA
\and
INAF--Osservatorio  Astronomico  di  Trieste,  via  G.B.  Tiepolo,  11, 34143 Trieste, Italy
\and
National  Optical  Astronomy  Observatory,  950  North  Cherry  Ave, Tucson, AZ 85719, USA
\and
Department of Astronomy, The University of Texas at Austin, Austin, TX 78712, USA
\and
Cavendish Astrophysics, University of Cambridge, Cambridge CB3 0HE, UK
\and
Kavli Institute for Cosmology, University of Cambridge, Cambridge CB3 0HE, UK}

   \date{Received ; accepted}
   
   \authorrunning{} \titlerunning{Lyman-$\alpha$ emitter fraction and galaxy physical properties at the edge of the epoch of cosmic reionization}
    
   \abstract{The fraction of Lyman-$\alpha$ emitters (LAEs) among the galaxy population has been found to increase from $z\sim0$ to $z\sim6$ and drop dramatically at $z>6$. This drop has been interpreted as an effect of an increasingly neutral intergalactic medium (IGM) with increasing redshift, while a Lyman continuum escape fraction evolving with redshift and/or a sudden change of galaxy physical properties can also contribute to the decreasing LAE fraction. We report the result of a large VLT/FORS2 program aiming to confirm spectroscopically a large galaxy sample at $z\geq6$ that has been selected in several independent fields through the Lyman Break technique. Combining those data with archival data, we create a large and homogeneous sample of $z\sim6$ galaxies ($N=127$), complete in terms of \lya\ detection at $>95\%$ for \lya\ equivalent width $\mathrm{EW}(\lya)\geq25\AA$. We use this sample to derive a new measurement of the LAE fraction at $z\sim6$ and derive the physical properties of these galaxies through spectral energy distribution (SED) fitting.
We find a median LAE fraction at $z\sim6$ lower than in previous studies, while our sample exhibits typical properties for $z\sim6$ galaxies in terms of UV luminosity and UV $\beta$ slope.  The comparison of galaxy physical properties between LAEs and non-LAEs is comparable to results at lower redshift: LAEs with the largest EW(\lya) exhibit bluer UV slopes, are slightly less massive and less star-forming. The main difference between LAEs and non-LAEs is that the latter are significantly dustier. Using predictions of our SED fitting code accounting for nebular emission, we find an effective  \lya\ escape fraction $\fefflya=0.23^{+0.36}_{-0.17}$ remarkably consistent with the value derived by comparing UV luminosity function with \lya\ luminosity function. We conclude that the drop in the LAE fraction from $z\sim6$ to $z>6$ is less dramatic than previously found  and the effect of an increasing IGM neutral fraction is possibly observed at $5<z<6$. The processes driving the escape of \lya\ photons at $z\sim6$ are similar to those at lower redshifts and based on our derived \fefflya, we find that the IGM has a relatively small impact on \lya\ photon visibility at $z\sim6$, with a lower limit for the IGM transmission to \lya\ photons, $\tigm\gtrsim0.20$, likely due to the presence of outflows.}


   \maketitle

   \section{Introduction}
   \label{sec:intro}

Cosmic reionization was a major phase transition in the early Universe history and large efforts have been made in the last decade to put constraints on when and how it occured, as well as identifying the main sources of ionizing photons.
Planck provided the most accurate measurement to date of the Thomson optical depth ($\tau= 0.066\pm0.013$), thus allowing to derive an instantaneous reionization redshift of $z=8.8\pm0.9$ \citep{Planck+16}. The currently leading candidate sources thought to be responsible for cosmic reionization are star-forming galaxies, with a main contribution coming from the faintest galaxies \citep[e.g.,][]{bouwens+15,bouwens+16a,finkelstein+15,robertson+15,livermore+17}, while the faint active galactic nuclei (AGN) contribution could be more important than previously thought \citep{madauhaardt15,giallongo+15}.

 In recent years, a growing number of galaxies with unambiguous ionizing photon leakage have been identified both in the nearby Universe and at high-redshift \citep{leitet+11,leitet+13,borthakur+14,debarros+16a,vanzella+16,shapley+16,izotov+16,bian+17}, but the total number of confirmed Lyman continuum (LyC) emitters remains small ($\lesssim10$), and this could be a consequence of view-angle effects with LyC photons escaping through a minority of solid angles \citep[e.g.,][]{kimmcen2014,cenkimm2015}. This lack of statistically significant samples of star-forming LyC emitters precludes the firm identification of ionizing photon leakage signatures necessary to identify LyC emitters into the reionization era, although some proposed diagnostics seem promising, such as the \oiii/\oii\ line ratio, the strength of interstellar absorption lines, a deficit of Balmer emission lines, or the structure of the \lya\ line \citep[e.g.,][but see also \citealt{rutkowski+17}]{heckman+11,jones+12,jaskotoey13,zackrisson+13,nakajimaouchi14,verhamme+15}. 

To circumvent the impossibility to directly observe ionizing photons escaping from high-redshift galaxies due to the intergalactic medium (IGM) opacity to LyC photons \citep[e.g.,][]{vanzella+15}, several spectroscopic surveys attempted to detect the \lya\ emission from star-forming galaxies at $z\geq6$, because resonant scattering in a partially neutral IGM will impact the detectability of \lya\ emission \citep{dayal+11}. Therefore the \lya\ photon visibility evolution can be used to put some constraints on cosmic reionization. The overall conclusion of these surveys is that a drop in the Lyman-$\alpha$ emitter (LAE) fraction among the Lyman Break galaxy (LBG) population is observed from $z\sim6$ to $z\sim7$ \citep[e.g.,][]{fontana+10,ouchi+10,stark+11,pentericci+11,ono+12,ota+12,caruana+14,schenker+14}, leading to the conclusion that cosmic reionization ended between $z\sim6$ and $z\sim7$. However, the amplitude of the median LAE fraction drop is such that it is difficult to explain by invoking plausible reionization models because it would require a extremely fast evolution of the neutral hydrogen fraction \citep{dijkstra+11,jensen+13}. \cite{dijkstra+14} suggested that this drop can be explained with a moderate increase of the neutral hydrogen fraction and an increasing Lyman continuum escape fraction with increasing redshift, reaching Lyman continuum escape fraction $\flyc\sim0.65$ at $z\sim6$, strongly contrasting with current constraints on \flyc\ for $z\sim3$ galaxies \citep[$\flyc<0.02-0.2$; e.g.,][]{vanzella+12,guaita+16,grazian+16}. However, alternative explanations have been proposed, such as an increase of the incidence of optically thick systems that would require lower neutral fraction \citep{boltonhaehnelt13}, or contribution from evolving galaxy properties \citep{mesinger+15}.

It has been pointed out in several studies that IGM could also strongly affect \lya\ visibility even in a fully ionized Universe. \cite{dijkstra+07} and \cite{zheng+10} derived at $z\sim6$ a mean IGM transmission to \lya\ photons ($\tigm\leq0.3$) and \cite{laursen+11} found $\tigm=0.26^{+0.13}_{-0.18}$. This low \lya\ transmission in a fully ionized Universe is due to the low transmission through the IGM of the \lya\ photons which are blueshifted because of their interaction with the ISM within galaxies \citep[e.g.,][]{verhamme+08}. At $z\sim3$, while some high-redshift LAEs exhibit double-peaked \lya\ emission \citep[e.g.,][]{vanzella+08}, the majority of them exhibit either weak or absent \lya\ blue bumps \citep{shapley+03,kulas+12}. Expanding shell models have been successful to reproduce observed \lya\ profiles \citep{verhamme+08} and study of interstellar absorption line velocities also supports a picture where outflows are ubiquitous in $z\sim3$ star-forming galaxies \citep{steidel+10}, while the IGM transmission to \lya\ photons is expected to be high. 

The \lya\ line can be used as a tool to constrain the IGM neutral fraction as long as the evolution of the galaxy physical properties influencing the escape of \lya\ photons is known. Indeed, at low- and intermediate redshift ($z\leq3$), \flya\ and EW(\lya) have been found to be related to the stellar mass (\mstar), star-formation rate (SFR), the age of the stellar population, ISM physical properties, and the dust extinction \citep[e.g.,][]{hayes+14,hathi+16,trainor+16}. Physical properties of $z\sim6$ galaxies can be derived directly from photometry, like the UV $\beta$ slope \cite[$f_\lambda\propto\lambda^\beta$; eg.,][]{bouwens+09,bouwens+12,bouwens+14,mclure+11,castellano+12,castellano+14,finkelstein+12,dunlop+13}, and through spectral energy distribution (SED) fitting \citep[e.g.,][]{eyles+05,eyles+07,schaererdebarros09,schaererdebarros10,mclure+11,curtislake+13,jiang+16}. Early analysis of $z\sim6$ photometry (combining optical, near-infrared and mid-infrared data) provided a picture where galaxies where already massive ($\mstar\geq10^{10}\msun$) and relatively old ($>100\mathrm{Myr}$) with a substantial Balmer break \citep{eyles+05,eyles+07,yan+05,yan+06}. \cite{zackrisson+08} showed that nebular emission and notably emission lines could have a large impact on photometry at high-redshift and this possibility has been explored in \cite{schaererdebarros09,schaererdebarros10}. Accounting for the impact of nebular emission on high-redshift galaxy physical properties is now a widespread approach
\citep[e.g.,][]{chary+05,robertson+10,robertson+13,vanzella+10a,vanzella+14,labbe+10,labbe+13,ono+12,oesch+13a,oesch+13b,oesch+14,stark+13,debarros+14,duncan+14,smit+14,salmon+15}. Taking into account nebular emission generally leads to lower stellar masses and younger ages, particularly for $z\sim6$ galaxies for which the two first {\it Spitzer}/IRAC bands are both contaminated by strong emission lines, namely \oiii+\hb\ and \ha\ respectively, while these two bands are providing the strongest constraints on both stellar mass and the Balmer break. Unfortunately, the fact that both bands IRAC1 and IRAC2 are contaminated prevents any empirical estimation of line contributions in those bands at $z\sim6$, unlike what can be done at $z\sim4$ \citep[e.g.,][]{shim+11,stark+13} and some other redshifts \citep{shivaei+15a,labbe+13,smit+14,faisst+16,marmol+16,rasappu+16}. Furthermore, the lack of constraints on galaxy redshifts also introduce uncertainties about which bands are going to be affected by emission lines.

In this paper, we present the analysis of a large sample of spectroscopically confirmed galaxies at $z\sim6$ from VLT/FORS2 observations  in 5 different fields \citep[][Pentericci et al., in prep.]{castellano+17}. Our analysis aims to derive the LAE fraction at $z\sim6$ and compare it with previous results to highlight which processes dominate the observed LAE fraction drop. We also derive physical parameters of this sample with SED fitting to check if the LAE fraction evolution is related to an evolution of the physical parameters instead of the evolution of the IGM neutral state. To increase the confidence on the derived physical parameters, while we minimize the number of assumptions going into our analysis, we check that the models reproduce observed properties but also reproduce predicted properties for which we do not have empirical constraints at $z\sim6$, namely \oiii+\hb\ emission line equivalent widths. We then study the physical properties, focusing on the relations between \lya\ emission and other physical parameters.

The paper is structured as follows. The selection procedure, spectroscopic and photometric data are described in Section 2. The results regarding the LAE fraction at $z\sim6$ are shown in Section 3, while we describe the SED fitting method and the derived physical properties in Section 4. In Section 5 we discuss the implications of our results regarding the IGM transmission to \lya\ photons at $z\sim6$. We summarize our conclusions in Section 6.

We adopt a $\Lambda$-CDM cosmological model with $\mathrm{H}_\mathrm{0}=70$ km s$^{-1}$ Mpc$^{-1}$, $\Omega_\mathrm{m} = 0.3$ and $\Omega_\Lambda= 0.7$. We assume a Salpeter IMF \citep{salpeter55}. All magnitudes are expressed in the AB system \citep{okegunn83}.

\begin{figure*}[htb]
\centering
\includegraphics[width=18cm,trim=0.25cm 0.25cm 0.25cm 0.25cm,clip=true]{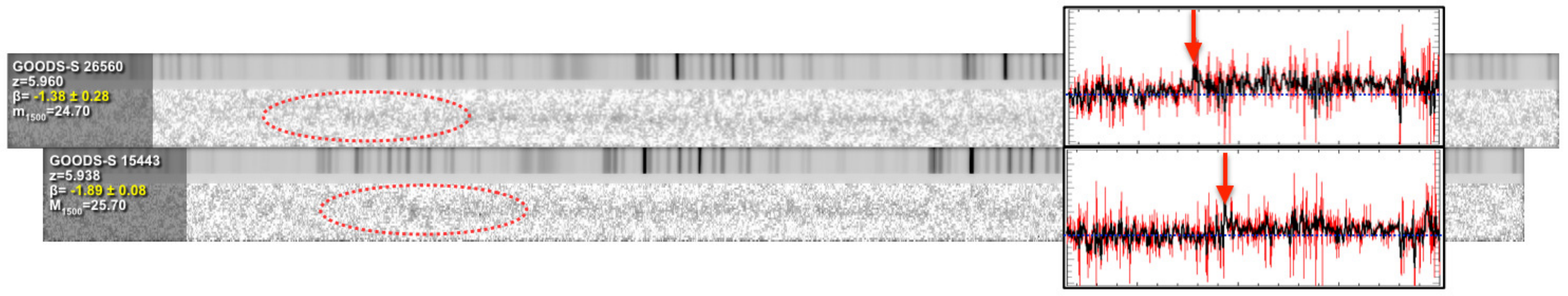}
\caption{Two-dimensional signal-to-noise spectra and sky spectra of two faint \lya\ emitters. The two insets show the corresponding 
one-dimensional FORS2 spectra with red arrows highlighting positions of the \lya-break, which are also visible in 2D spectra. For the galaxy on the top, $\mathrm{EW(\lya)\leq1.3\AA}$ and for the galaxy on the bottom $\mathrm{EW}(\lya)=10.0\pm1.3\AA$.}
\label{fig:fig1}
\end{figure*}

\begin{figure}[htb]
\centering
\includegraphics[width=8.75cm,trim=1cm 0cm 1.7cm 1cm,clip=true]{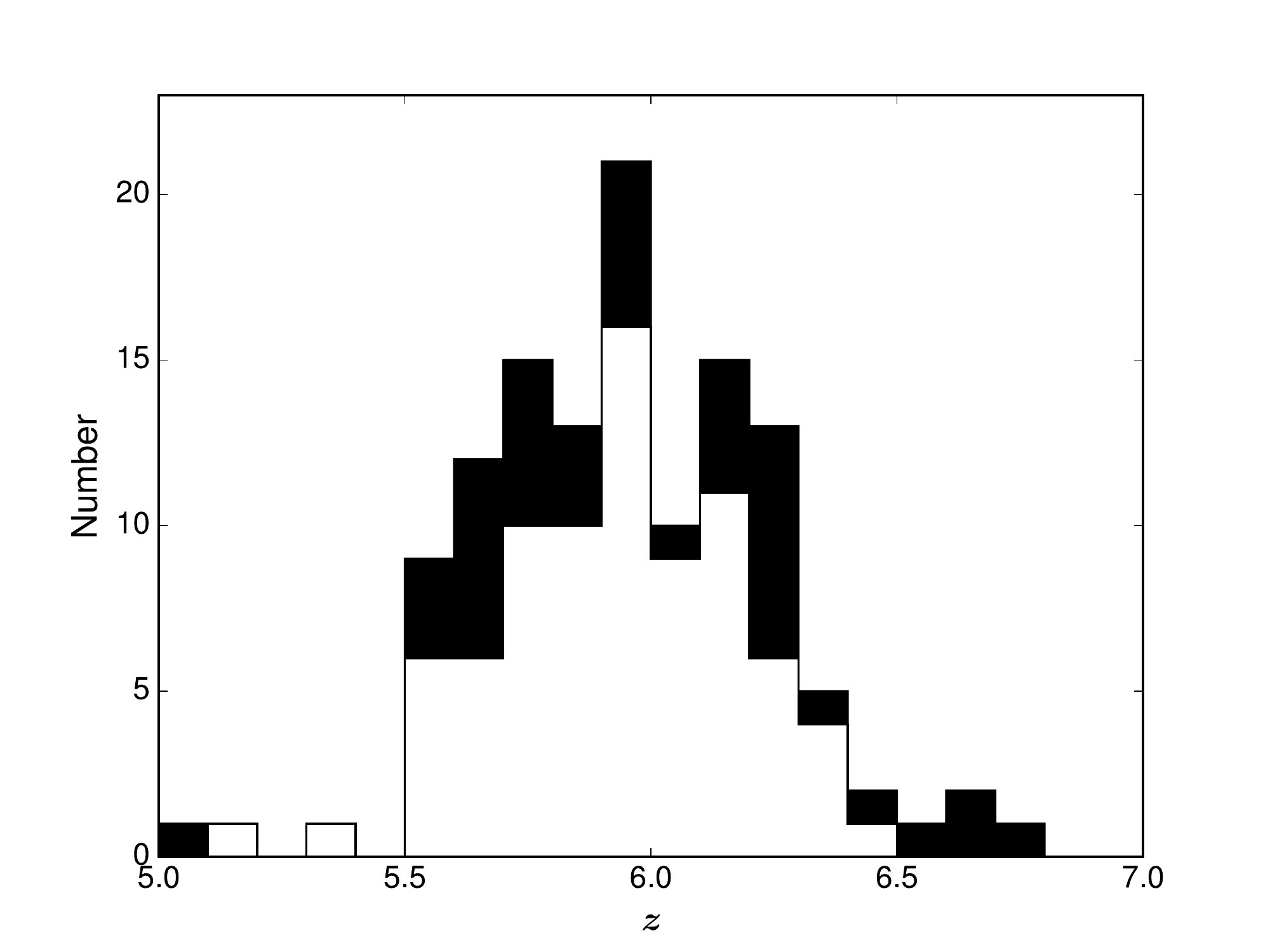}
\caption{Redshift distributions for the spectroscopically confirmed and photometric samples in white and black, respectively.}
\label{fig:fig2}
\end{figure}

   \section{Data}
   \label{sec:data}
   
   \subsection{Spectroscopic data}
   \label{sec:spectro}

We use  data obtained in the context of CANDELSz7 and ESO large program  (ID: 190.A-0685, PI: L. Pentericci) that acquired deep observations of three of the CANDELS fields: the Great Observatories Origins Deep Survey \citep[GOODS,][]{giavalisco+04b} Southern field, the Cosmological Evolution Survey \citep[COSMOS,][]{scoville+07,koekemoer+07} field, and the UKIDSS Ultra-Deep Survey \citep[UDS,][]{lawrence+07,cirasuolo+07,mortlock+15} field. 140 hours observation have been allocated to this program. We also add the results from other VLT/FORS2 program \citep[ID: 085.A-0844, 084.A-0951, 088.A-0192, PI: A. Fontana,][]{fontana+10,castellano+10a,castellano+10b} with a total of 63 hours observation distributed among the New Technology Telescope Deep Field \citep[NTTDF,][]{arnouts+99,fontana+00,fontana+03}, the Bremer Deep Field \citep[BDF,][]{lehnertbremer03}, the GOODS-S and UDS fields. Finally, we also use the results from an archival VLT/FORS2 program \citep[ID: 088.A-1013, PI: A. Bunker;][]{caruana+14} which targeted the Hubble Ultra Deep Field (HUDF) with 27 hours allocated. All the data combined provide typically 15-30 hours integration time for each target. The FORS2 600z grism configuration provides a useful area of $7'\times4.33'$. Eight masks were used in the CANDELSz7 program, with five additional masks in total from the other programs.

\begin{figure}[htb]
\centering
\includegraphics[width=8.75cm,trim=0.75cm 0cm 1.5cm 1cm,clip=true]{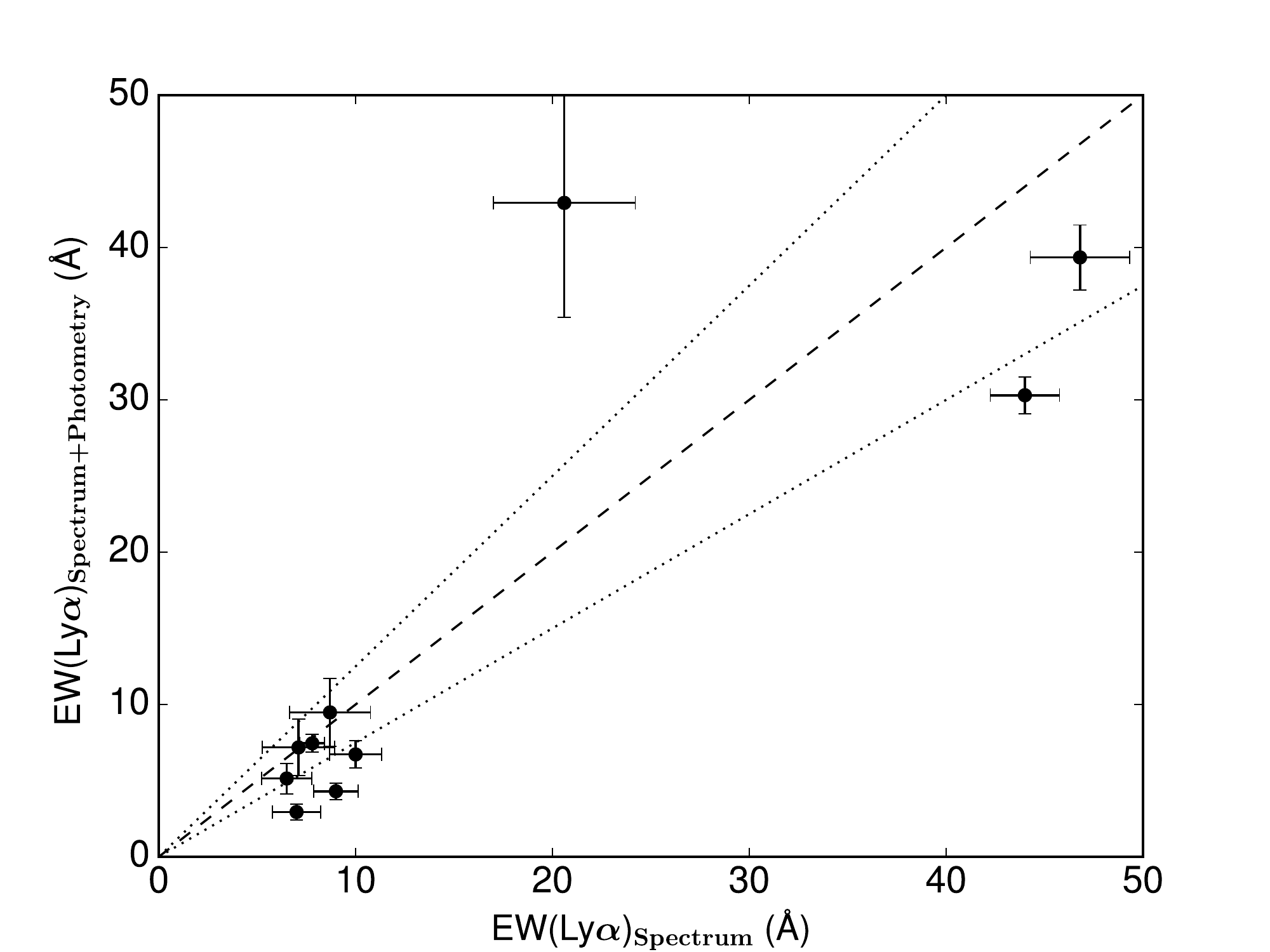}
\caption{Comparison of EW(\lya) derived from spectra only and from spectra and photometry (see Sec.~\ref{sec:ewlya}). We show the results for the 10 galaxies where the continuum is spectroscopically detected. The dashed line shows the one to one relation and dotted lines show $\pm25\%$ from the one to one relation.}
\label{fig:fig3}
\end{figure}

For the GOODS-S, COSMOS and UDS fields, we use the publicly available CANDELS \citep{grogin+11,koekemoer+11} catalogs for each of those fields \citep[][respectively]{guo+13,nayyeri+17,galametz+13} with a wavelength coverage from $\sim0.3\micron$ to 8.0\micron. For the BDF and NTTDF fields, we use the available photometry in the $V$, $R$, $I$, $Z$, $Y$, $J$, and $K_S$ bands \citep{castellano+10b}. Details of the observation and data reduction are lengthly described in the different papers cited previously, we refer the reader to those references as well as Pentericci et al. (in prep.) for the CANDELSz7 survey description.

\begin{figure}[htb]
\centering
\includegraphics[width=8.75cm,trim=1cm 0cm 2cm 1cm,clip=true]{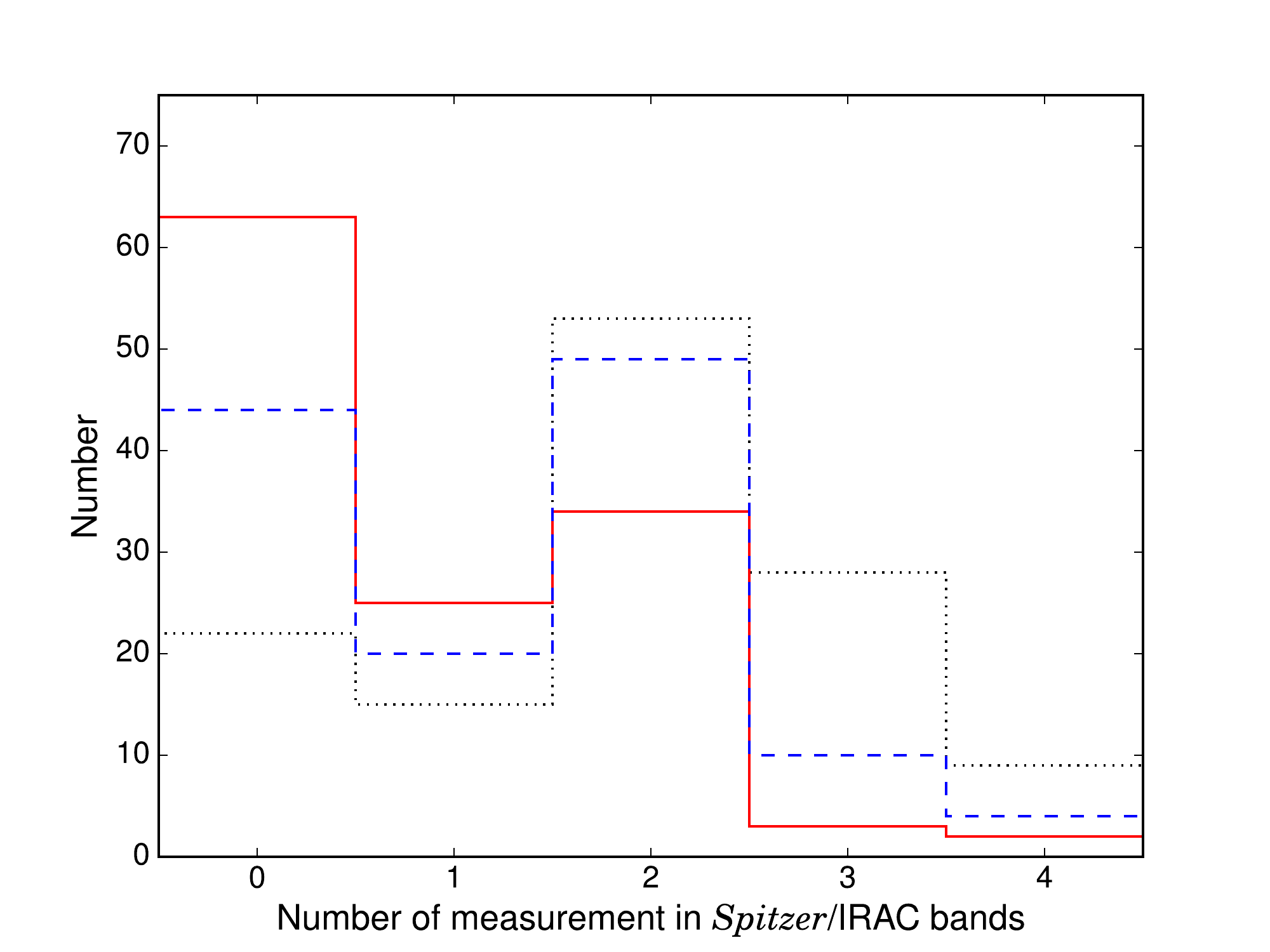}
\caption{Distribution of the sample as a function of the number of measured fluxes in {\it Spitzer}/IRAC bands (black histogram). We show in blue and red the distribution of the flux measurement with $S/N\geq2$ and $S/N\geq3$, respectively.}
\label{fig:fig4}
\end{figure}

The targets from CANDELS fields have been homogeneously selected using a selection similar to the one used in \cite{bouwens+15b}, with the $i$-dropout criteria being:
\begin{eqnarray}
i_{775}-z_{850}>1.0;\nonumber\\Y_{105}-H_{160}<0.5;\nonumber\\(S/N(B_{435})<2)\land (V_{606}-z_{850}>2.7\lor S/N(V_{606})<2)
\label{eq:dropout}
\end{eqnarray}
to which we add an additional criterion with $H_{160}<27.5$. For the BDF and NTTDF, given that no \hst\ data are available, the detection band was the HAWK-I $Y$-band. Although this is bluer than the CANDELS $H$-band, it is still free from contamination from emission line at $z\sim6$. For BDF and NTTDF, selection criteria are:
\begin{eqnarray}
I-Z>1.3;\nonumber\\ S/N<2\ \textrm{in all bands blueward of }I\textrm{-band}
\label{eq:dropout2}
\end{eqnarray}
with $Y<26.5$ for the two fields. While the criteria are slightly different for BDF and NTTDF compared to the other fields, the color criteria is equivalent. Galaxies from these two fields make only a small fraction of our entire sample (9\%). The only two spectroscopically confirmed low-redshift contaminants have been observed in the NTT and BDF fields ($z=1.3$ and $z=0.5$, respectively).

The VLT/FORS2 wavelength coverage ranges from 5700\AA\ to 10000\AA\ (depending on the slit position), allowing the detection of \lya\ emission at $z\sim4-7.2$ if present, and even in some cases detect directly the continuum emission and the Lyman break. Furthermore, the wavelength coverage allows us to identify low-redshift interlopers by detecting multiple emission lines, e.g., \oiiidoub, \ha, or \oiilam\ emissions at $z\leq1$, $z\leq0.5$, and 0$.5\leq z\leq1.7$, respectively \citep[e.g.][]{vanzella+11}. In case of single emission detection, for $S/N>6$ the resolution $R=1390$ allows us to identify the typical asymmetric profile of \lya, which can be used to discriminate between low- and high-redshift galaxies. We also check for inconsistency between the measured spectroscopic redshift and the redshift probability distribution function. Data reduction is performed as described in \cite{vanzella+11,vanzella+14b} following the ``A-B'' dithering scheme method.  The high-quality of the deep spectra allows us to identify faint $z\sim6$ \lya\ emission as shown in Fig.~\ref{fig:fig1}. 

The final sample consists of 127 galaxies, with 74 galaxies in GOODS-S (58\% of the sample), 23 in UDS (18\%), 19 in COSMOS (15\%), 7 in BDF (6\%), and 4 in NTT (3\%). For 81 of these galaxies, we can derive a spectroscopic redshift, mainly from the presence of the \lya\ emission line or in few bright cases ($N=10$), from the presence of continuum emission with a drop consistent with the Lyman break (Fig.~\ref{fig:fig1}). All objects with a spectroscopic confirmation will have their properties published in a Table in Pentericci et al. (2017, in prep.). For 46 objects, it was not possible to derive a spectroscopic redshift given that no features were detected in the deep spectra. In this paper, we assume that all the undetected objects are also at $z\simeq6$, consistently with their photometric redshifts (Fig.~\ref{fig:fig2}). Assuming that all undetected objects are low-z interlopers and accounting for the two spectroscopically confirmed low-z galaxies in NTT and BDF, we can set a robust upper limit for the interloper fraction with 38 interlopers over a sample of 129 observed objects. The upper limit for the interloper fraction is then $\leq29\%$.


\subsection{EW(\lya) measurement}
\label{sec:ewlya}

\lya\ equivalent widths are derived by using the nearest redward \hst\ bands from the \lya\ emission (excluding bands affected by \lya) and deriving the UV $\beta$ slope directly from the photometry \citep[][Pentericci et al. in prep.]{castellano+12}. To derive the UV $\beta$ slopes, we derived fluxes using apertures of 1.75FWHM instead of isophotal as in the CANDELS published catalogs. Indeed, we found that for small galaxies as in our sample, colors measured in small apertures are more stable than the original isophotal ones. To estimate the continuum at 1216\AA\ (rest-frame), we used the $Y-$ or $J$-band magnitudes from the CANDELS catalogs that are corrected to total flux via aperture correction. For non detection of the \lya\ line, 3$\sigma$ upper limits on the equivalent width have been derived from the S/N of \lya\ lines as described in \cite{vanzella+14b}.
We do not apply any aperture correction to the \lya\ flux measurements. Even if there is extended \lya\ emission due to the scattering of \lya\ photons in the  circum galactic medium \citep[CGM,][]{wisotzki+16}, the relatively small intrinsic sizes of the observed galaxies \citep{curtislake+16} should prevent flux losses. Furthermore, no correction is usually applied at $z\sim6$ \citep[e.g.,][]{stark+11}. The error on the $\beta$ slope is not propagated to the EW(\lya) estimation, but we estimate that the total error on EW(\lya) is not larger than 35\%. We show in Fig.~\ref{fig:fig3} the comparison between EW(\lya) derived from spectra only for the 10 galaxies for which the continuum is spectroscopically detected, and using the aforementioned method. While this small subsample is strongly biased toward bright galaxies and so toward galaxies with faint \lya\ emission (Sec.~\ref{sec:lyauv}), EW(\lya) measurements from the two methods are consistent.

\begin{figure*}[htb]
\centering
\includegraphics[width=18cm,trim=1.25cm 0cm 2cm 1.15cm,clip=true]{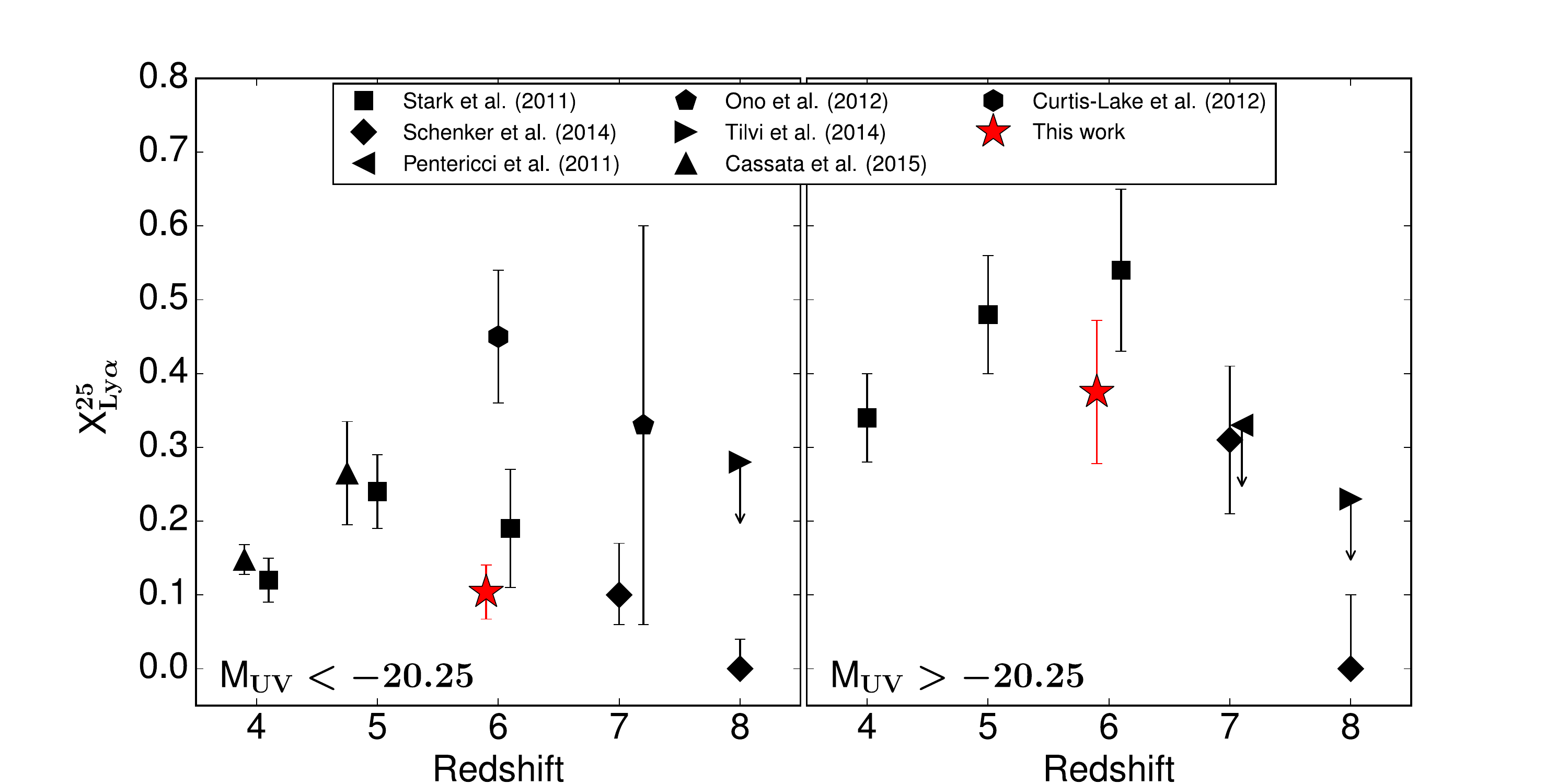}
\caption{Fraction of LAEs (with $\mathrm{EW}(\lya)>25$\AA, X$_\mathrm{\lya}^{25}$) at $4\leq z\leq8$ for the brightest ($\mathrm{M}_\mathrm{UV}<-20.25$, left panel) and faintest ($\mathrm{M}_\mathrm{UV}>-20.25$, right panel) galaxies. We show the results from \cite{stark+11}, \cite{schenker+14}, \cite{ono+12}, \cite{pentericci+11}, \cite{tilvi+14}, \cite{cassata+15}, and \cite{curtislake+12}. We introduce slight offsets in redshift to increase clarity when necessary.}
\label{fig:fig5}
\end{figure*}

\begin{figure}[htb]
\centering
\includegraphics[width=8.75cm,trim=0cm 0cm 0cm 0cm,clip=true]{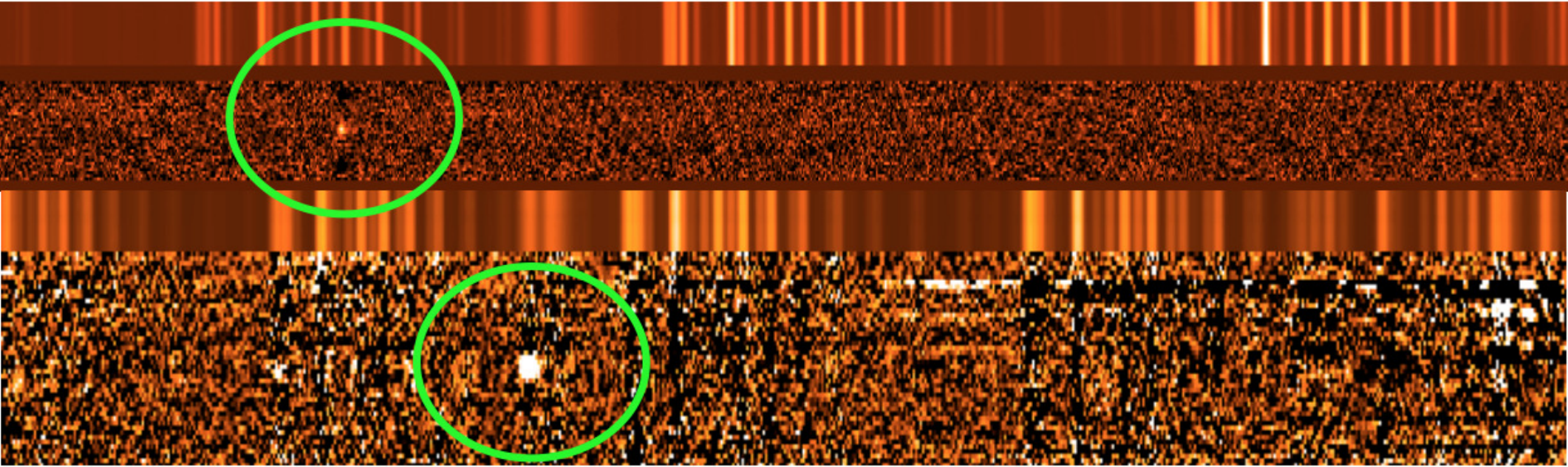}
\caption{2D signal-to-noise and sky spectra for two faint LAEs at $z=5.92$ (top) and $z=6.09$ (bottom) with $\mathrm{EW}(\lya)=101\AA$ and $\mathrm{EW}(\lya)=330\AA$, respectively (see text).}
\label{fig:fig6}
\end{figure}   

For all the programs the slit width was always 1''. For Bunker's program some slits had a slit length as low as 6'', the minimum slit length was 8'' for the CANDELSz7 program, and the slit length was always above 10'' for Fontana's program. Simulations have been performed to estimate the minimum line flux that is measurable for this sample. These simulations are similar to those performed in \cite{vanzella+11,vanzella+14} and \cite{pentericci+14}. Artificial two-dimensional asymmetric \lya\ lines have been added in the raw science frame, exploring a range of typically observed fluxes and full width half-maxima (FWHM; varying emerging values from 280 to 520 km s$^{-1}$), with a fiducial FWHM of 300 km s$^{-1}$. We then apply the reduction pipeline and response curve. At fixed flux, the larger the FWHM is, the lower the S/N is. Going from 280 to 520 km s$^{-1}$ (emerging FWHM), the S/N decreases by a factor of $\sim1.5$.  We derive that our observations detect \lya\ flux as low as $2.2\times10^{-18}$ erg s$^{-1}$ cm$^{-2}$ at 3$\sigma$ over the entire wavelength range probed in this work (which means that we can recover even fainter \lya\ lines in wavelength ranges free of sky lines). Assuming that the nearest band redward of \lya\ (and not contaminated by this line) provides a measurement of the continuum at the \lya\ wavelength, we estimate that our sample is complete at $>95\%$ for \lya\ equivalent width of $\mathrm{EW}(\lya)\geq25\AA$.

Due to the importance of the {\it Spitzer}/IRAC detections to constrain the stellar mass, the age (Balmer break), and emission line contribution \citep[e.g.,][]{jiang+16}, we show in Fig.~\ref{fig:fig4} the number of flux measurements in {\it Spitzer}/IRAC bands.


\section{A lower \lya\ emitter fraction at $z\sim6$}

   Thanks to our large spectroscopic sample we can derive the LAE fraction (defined as the fraction of LAE with rest frame equivalent width EW(\lya)$>25$\AA) at $z\sim6$ which allows us to put constraints on the ionization state of the IGM. This fraction has already been described as rising from $z\sim4$ to $z\sim6$ \citep[][S11 hereafter]{stark+11} and rapidly declining at $z>6$ \citep[e.g.,][]{pentericci+11}.  This evolution of the LAE fraction between $z\sim6$ and $z>6$ has been interpreted as the effect of partially neutral IGM on the \lya\ photons emitted from high-z galaxies \citep[e.g.,][]{schenker+12} and so as our witnessing of the cosmic reionization end.
   
         \begin{figure}[htb]
\centering
\includegraphics[width=8.75cm,trim=1cm 3.25cm 1.75cm 4.25cm,clip=true]{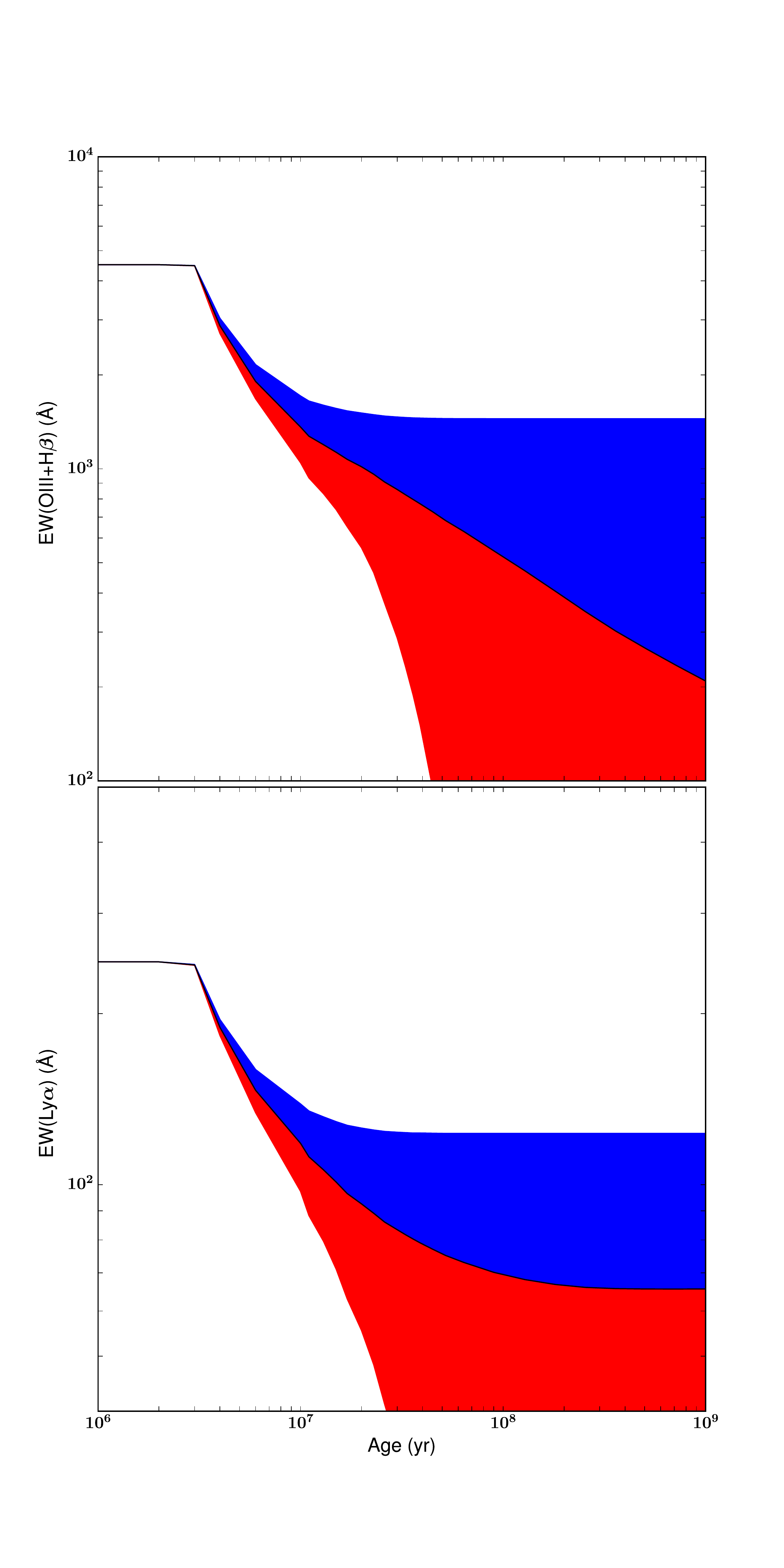}
\caption{Top: EW(\oiii+\hb) evolution with age for $Z=0.2Z_\odot$ and the range of star formation used in this work: exponentially rising (blue), and exponentially declining (red). Bottom: same for EW(\lya).}
\label{fig:fig7}
\end{figure}   
   
   We show in Figure~\ref{fig:fig5} the LAE fractions for our bright ($\mathrm{M}_\mathrm{UV}<-20.25$) and faint ($\mathrm{M}_\mathrm{UV}>-20.25$) subsamples. The absolute UV magnitude M$_\mathrm{UV}$ refers to the absolute magnitude at 1500\AA. To determine it for each galaxy, we use the integrated SED flux in an artificial filter of 200\AA\ width centered on 1500\AA. Our data covers 5 fields which should mitigate the cosmic variance effect on our results \citep[e.g.,][]{ouchi+08}.  Comparing our results with previous studies \citep[e.g.,][]{stark+11,pentericci+11}, the LAE fractions that we derived at $z\sim6$ are consistent with a trend of increasing and then decreasing LAE fraction with redshift, the main uncertainty being at which redshift is the peak of the LAE fraction. However, the LAE fraction found by \citet[hereafter CL12]{curtislake+12} at $z\sim6$ is significantly higher compared to our result or S11. We check that the color selection criterion $i-z$ is not the cause of this discrepancy by applying the same criterion as S11 to our sample ($i-z>1.3$), as well as the criterion ($i-z>1.7$) used in CL12, in both cases to the bright sample ($\muv<-20.25$). While the fraction of LAE (X$_\mathrm{\lya}^{25}$) increases (up to 0.16 using the CL12 criterion), the fraction of unconfirmed sources (defined as sources for which we cannot assign a spectroscopic redshift) remains constant $\sim0.35$ for all $i-z$ cut, showing that interlopers are unlikely to be the cause of the difference between the results of CL12 and our own (and those of S11). \lya\ equivalent widths are derived using a narrow-band filter in CL12, instead of a broad-band filter in S11/this work, but it is also unlikely that this difference can introduce such a large discrepancy because the maximum 3$\sigma$ EW(\lya) upper limit for unconfirmed object is $<16\AA$ and there is only one object with a measured EW(\lya) near the EW threshold used ($25\AA$). Therefore, it would require a difference in the UV continuum flux estimation by a large factor ($>2$) to explain the difference between our work and CL12. Alternative explanations can be the small statistics of the CL12 sample or the fact that particularly bright LBGs can exhibit unusually strong \lya\ emission \citep[e.g.,][]{matthee+17}.
   
   Comparing our results with S11, we find a lower median LAE fraction in both the bright and faint samples, and while results are consistent within 1$\sigma$ uncertainties between the two studies, we still discuss the differences between the medians as these values have been used to study galaxy evolution and cosmic reionization \citep[e.g.,][]{dijkstra+14}.

    Our sample size is $\sim60\%$ larger than the S11 sample. Furthermore, at $z\sim6$ the \lya\ line can affect the dropout selection \citep{stanway+08}: for a sample relying only on the $z_{850}$ band as the detection band, strong emitters are scattered in this band because of the emission line strength, and not because of the continuum.
	The main difference in terms of color criteria comes from our IR criterion, i.e., $H<27.5$. Without this IR criterion, color selection of $z\sim6$ is biased toward faint galaxies with strong \lya\ emission: we show in Fig.~\ref{fig:fig6} two examples of galaxy with large EW(\lya) ($>100\AA$) and with $z_{850}=27.50$ (top) and $z_{850}=26.30$ (bottom). Those two galaxies would have been selected using S11 color criterion while they are not in our sample because they are not detected in $H_{160}$. Once corrected for the \lya\ contribution, their $z_{850}$ magnitudes are $>28.28$ and $>27.40$, respectively, which implies that {\it non}-\lya\ emitters as faint as those two LAEs are likely absent from the S11 sample. Our selection criteria should prevent such bias and at a given luminosity we should select both LAE and non-LAE. We conclude that the most likely reason for the differences between our results on LAE fraction and S11/CL12 is due to the fact that our sample is $H$-band detected, while other samples are $z$-band detected, with the $z$-band being contaminated by the \lya\ emission.
	
	 Considering our upper limit on interlopers ($\leq29\%$), our measured LAE fractions can be considered as lower limits and the true value should be between 1 and the values given in Fig.~\ref{fig:fig5}. However, our upper limit has been derived assuming that all undetected objects are interlopers, which we consider as unlikely given the typical exposure time (Sec.~\ref{sec:spectro}). Soon, \jwst\ should permit spectroscopic redshift confirmation even for galaxies with weak or absent \lya.
	 
In \cite{dijkstra+14} different models of cosmic reionization are explored to derive the EW(\lya) cumulative distribution at $z\sim7$ from the cumulative distribution at $z\sim6$, assuming a fully ionized Universe at $z\sim6$ (see their Fig.~3). Matching the $z\sim7$ distribution required an extremely rapid evolution of the neutral fraction $\Delta x_{\hi}\sim0.5$ and they explore a scenario where the Lyman continuum escape fraction increases with redshift, alleviating the requirement for a rapid evolution of the IGM state.  Using our sample with $\mathrm{M}_\mathrm{UV}>-20.25$ and EW(\lya) measurements lead to a EW(\lya) cumulative distribution with a steeper slope than in \cite{dijkstra+14}, with values $\mathrm{P(EW>25\AA)}\sim0.5$, $\mathrm{P(EW>50\AA)}\sim0.2$, and $\mathrm{P(EW>75\AA)}\sim0.1$. This latter value is consistent with the upper limit on $\mathrm{P(EW>75\AA)}$ at $z\sim7$ used in \cite{dijkstra+14}. A detailed analysis, including a comparison between the $z\sim6$ and $z\sim7$ \lya\ equivalent distribution derived from our new survey will be presented in a forthcoming paper.

Based on our results on the EW(\lya) distribution and on the evolution of the LAE fraction (Fig.~\ref{fig:fig5}), we conclude that the IGM evolution from $z\sim6$ to $z\sim7$ is less dramatic than previously thought, assuming that the LAE visibility mostly depends on the IGM state. The median LAE fraction plus 1$\sigma$ uncertainties that we derive at $z\sim6$ is also consistent with a flattening of the relation between LAE fraction and redshift at $5<z<6$, which can imply that the IGM neutral fraction starts to increase between $z\sim5$ and $z\sim6$, more likely at $5.5<z<6.0$ \citep{becker+15}. We study in details the galaxy physical properties of our sample to check if they differ from the properties of other samples and if those properties can affect the derived LAE fraction.

\begin{figure}[htb]
\centering
\includegraphics[width=8.75cm,trim=0.25cm 0cm 1.5cm 0.75cm,clip=true]{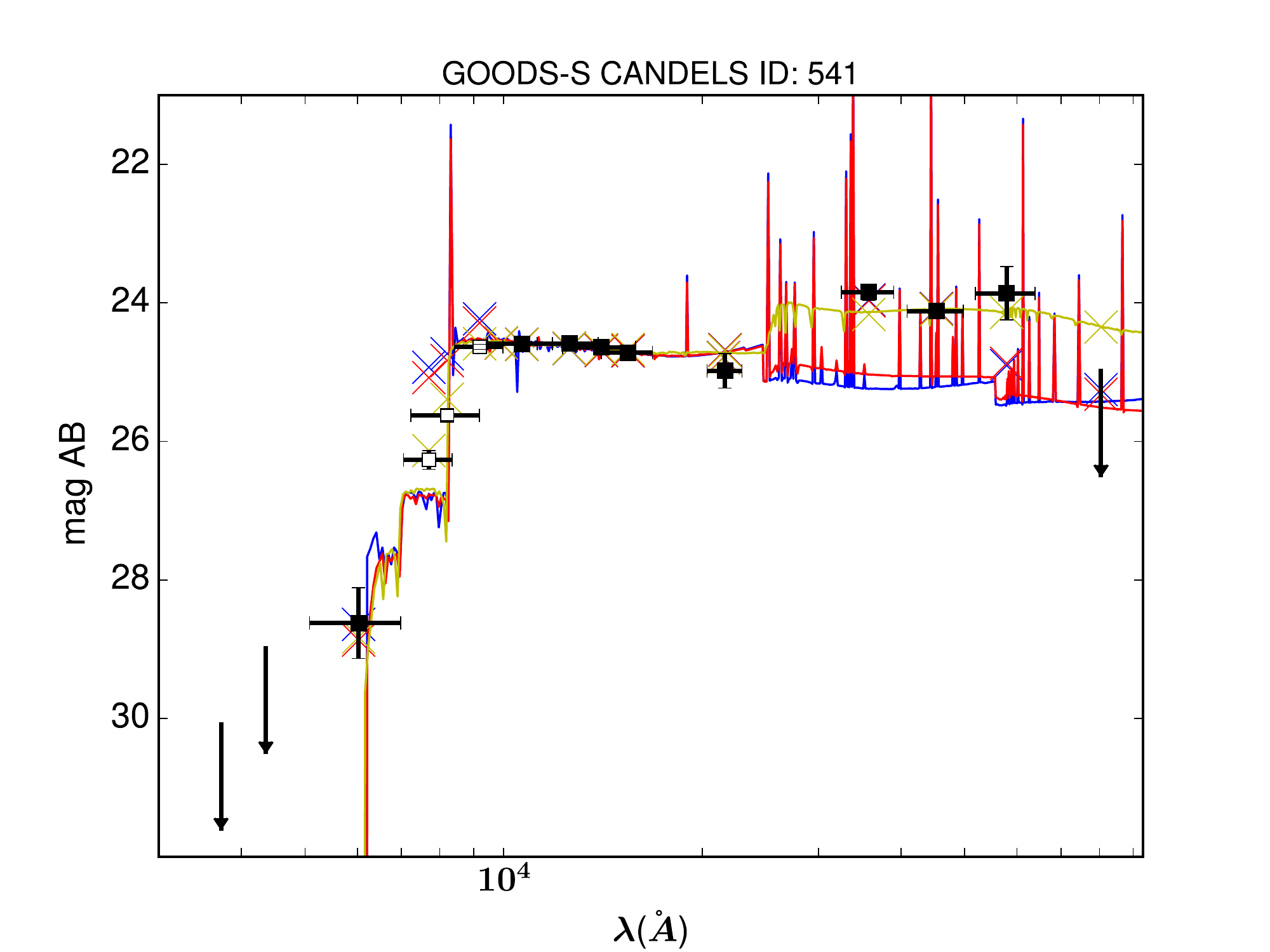}
\caption{Observed (black and white squares) and best-fit SEDs (solid lines) of a $z=5.786$ galaxy. White squares show bands not used in the SED fitting because of possible impact of the \lya\ line. The errorbars of the observed wavelength indicate the surface of the normalised filter transmission curve. SED fits in blue and red are based on \cite{BC03} models with (red) and without (yellow) nebular emission, and a fit using BPASSv2.0 \citep{eldridgestanway16,stanway+16} is shown in blue (Sec.~\ref{sec:bpass}). Color crosses show the synthesised flux in the filters. Without nebular emission age and stellar mass are $\sim10^9\mathrm{yr}$ and $\sim10^{10.5}\msun$ respectively, while accounting for nebular emission gives $\sim10^6\mathrm{yr}$ and $\sim10^9\msun$.}
\label{fig:fig8}
\end{figure}  

\section{Galaxy physical properties}

While the LAE fraction evolution is used to constrain the IGM neutral fraction, it is known that the \lya\ properties are related to other physical properties like luminosity and UV slopes, and the increase of the LAE fraction up to $z\sim5-6$ is likely related to the decreasing galaxy dust obscuration with increasing redshift \citep[e.g.,][]{stark+10}. In the following, we compare the \lya\ properties of our sample with other physical properties to assess which parameters are affecting the LAE fraction.

   \subsection{SED modeling}
   \label{sec:sed}
     
   To derive the galaxy physical properties such as stellar mass or age of the stellar population, we use a modified version of \hyperz\ \citep{bolzonella+00}, accounting for nebular emission \citep[lines and continuum,][]{schaererdebarros09,schaererdebarros10}. We generate a set of spectral templates with the GALAXEV code of \cite{BC03}, for three different metallicities (Z=0.004, 0.04, 0.02) and using a unique star-formation history (SFH) defined with a star-formation rate as $\mathrm{SFR}\propto\exp(t/\tau)$ with $\tau=[-10, -30, -50, -70, -100, -300, -500, -700, -1000, -3000,\\ \infty, 3000, 1000, 700, 500, 300, 100, 70, 50, 30, 10]$Myr. 
This choice of SFH also allows us to explore a large range of possible emission line equivalent widths (Fig~\ref{fig:fig7}). The stellar age is defined as the age since the onset of star-formation. We do not consider a minimal age because the dynamical timescale at $z\sim6$ is likely to be low \citep[$t_d\sim10\mathrm{Myr}$, e.g.,][]{debarros+14}.
      
       \begin{figure}[htb]
\centering
\includegraphics[width=8.75cm,trim=1cm 0cm 2cm 1.25cm,clip=true]{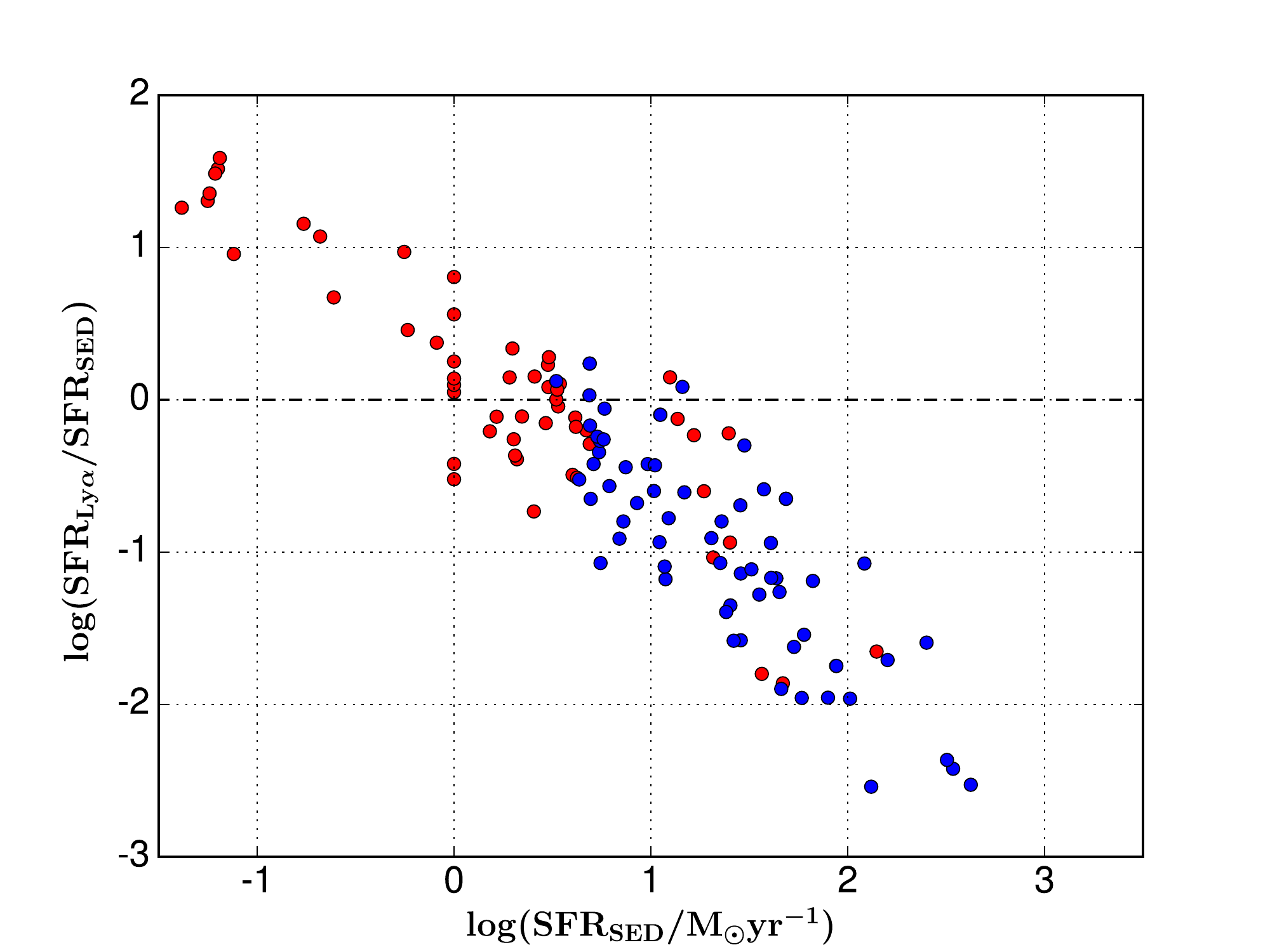}
\caption{Relation between $\log(\sfrsed)$ and $\log(\sfrlya/\sfrsed)$ for galaxies with detected \lya\ emission, assuming a declining (red dots) and a constant SFH (blue dots). \lya\ luminosities have not been corrected for dust. The dashed line shows $\log(\sfrlya)=\log(\sfrsed)$. We remind that \sfrlya\ is a lower limit to the true SFR (see text).}
\label{fig:fig9}
\end{figure}
      
   We consider three different dust attenuation curves: the SMC curve \citep{prevot+84}, the Calzetti curve \citep{calzetti+00}, and the Reddy curve \citep{reddy+15}. The Calzetti and Reddy curves have been derived using Balmer decrement on local and $z\sim2$ samples of star-forming galaxies, respectively, while there is mounting evidence that at $z\sim6$ the most appropriate curve is an SMC-like curve \citep{capak+15,bouwens+16b}, and it could be also the case at $z\sim2$ \citep{reddy+17}. For simplicity, we use the same dust attenuation curve for both the stellar continuum and the nebular emission, while using a Calzetti curve with this assumption yields SFR(SED) consistent with SFR(UV+IR) \citep{shivaei+15a}. A typical ratio has been generally assumed between nebular and stellar color excesses \citep[][]{calzetti+00}, but recent observations have shown that at $z\sim2$ there is no simple linear relation between nebular and color excesses, while the average ratio is $\sim1$ \citep{reddy+15}. For simplicity, we assume that $\mathrm{E(B-V)}_\mathrm{stellar}=\mathrm{E(B-V)}_\mathrm{nebular}$. We exclude bands possibly affected by the \lya\ emission from the SED fitting.

Minimization of $\chi^2$ over the entire parameter space yields the best-fit SED. Best-fit parameters are assumed to be the median of the marginalized likelihood and uncertainties are determined through the likelihood marginalization for each parameter of interest with $\cal{L}\propto\exp$$(-\chi^2/2)$.

      \begin{figure}[htb]
\centering
\includegraphics[width=8.75cm,trim=0.5cm 0cm 1.5cm 1cm,clip=true]{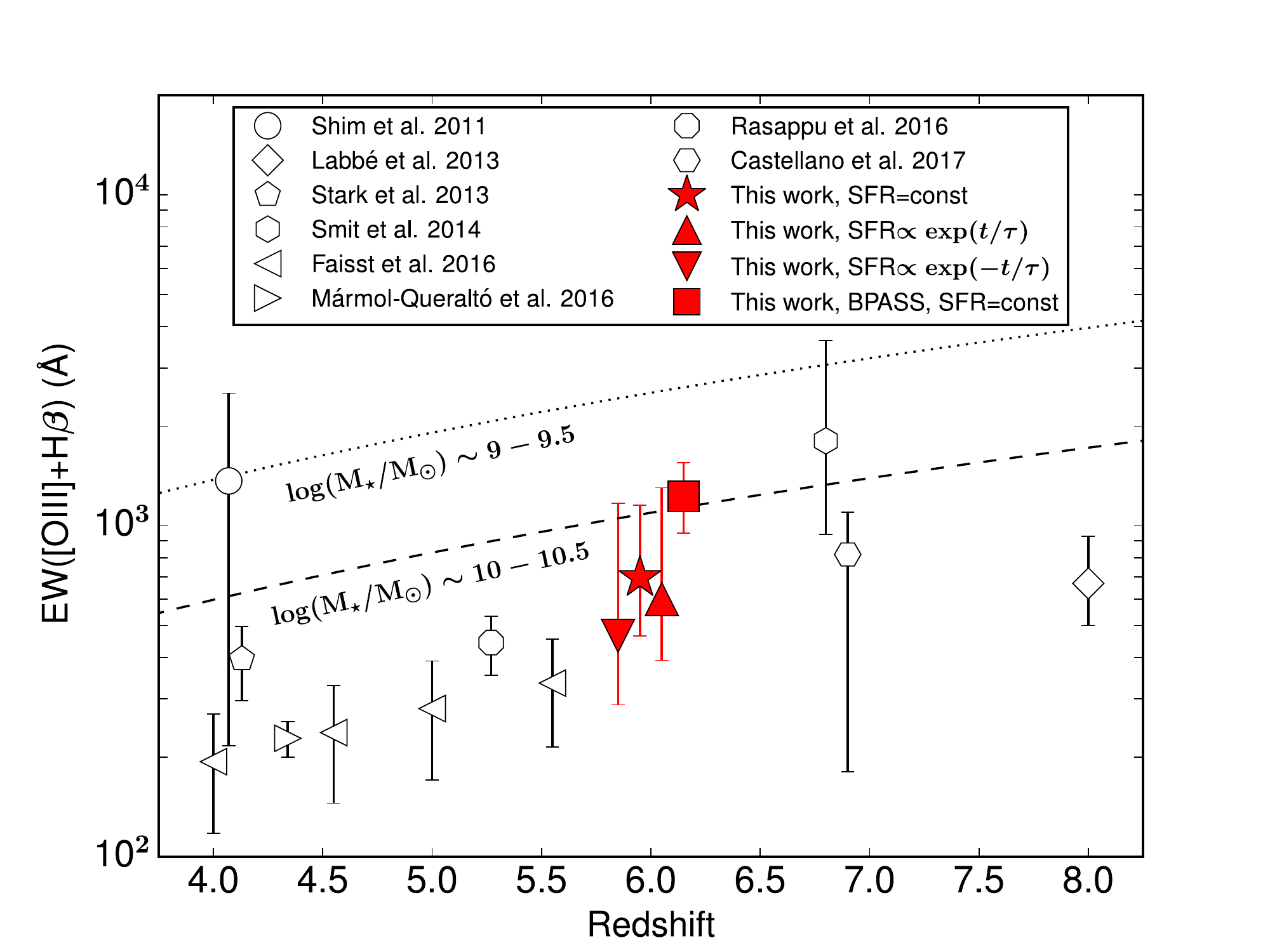}
\caption{EW(\oiii+\hb) vs. redshift for different studies deriving empirically EW from the photometry \citep{shim+11,labbe+13,stark+13,smit+14,faisst+16,marmol+16,rasappu+16,castellano+17} and the results obtained for the three star-formation histories used in the present work (rising, constant, and declining) assuming a SMC attenuation curve (alternative curves do not affect the results) and BC03 templates. We also show the result using BPASS templates (Sec.~\ref{sec:bpass}). For the studies deriving EW(\ha) we assume the typical ratio between this line and \oiii+\hb\ from \cite{AF03} for a metallicity $Z=0.2Z_\odot$. We show the relation between EW and redshift derived in \cite{fumagalli+12}, extrapolated to high-redshift and for two different stellar masses \citep{smit+14}.}
\label{fig:fig10}
\end{figure}

While we cannot compare \sfrsed\ with usual SFR tracers like \sfrha\ or \sfruvir, we can derive \sfrlya\ \citep{atek+14}, keeping in mind that the \sfrlya\ is always providing a firm {\it lower limit} of the true SFR because of the loss of \lya\ photons due to the radiative transfert effect of the interstellar medium and dust attenuation \citep[ISM,][]{schaererverhamme08,verhamme+08}, the possible effect of the intergalactic medium \citep{zheng+10,dijkstra+11,laursen+11}, and also slit loss. By comparing \sfrlya\ and \sfrsed\ (Fig.~\ref{fig:fig9}), we show that assuming a declining SFH leads to an underestimate of the true SFR for $\sim50\%$ of the sample for which we are able to measure a \lya\ flux.
The SFR underestimation under the assumption of a declining SFH is similar to results obtained at $z\sim2$ \citep{wuyts+11,reddy+12a,price+14}. Assuming a constant or rising SFH leads to  \sfrsed\ consistent with \sfrlya, therefore in the following, we exclude declining SFHs from the range of possible SFH.
We note that in the range of stellar mass explored with our $z\sim6$ sample, effects of stochastic star-formation history are expected \citep[e.g.,][]{hopkins+14}. These can lead to difference in SFR estimation because of the different timescales probed by different tracers \citep[e.g., UV and nebular emission lines;][]{dominguez+15}. While we explore a large range of SFH and ages younger than 100 Myr \citep[the UV to SFR conversion assumes a constant SFH for 100 Myr;][]{kennicutt98}, which make possible large differences between \sfruv\ and \sfrsed\, we find that on average \sfrsed\ and \sfruv\ do not differ by more than a factor of 2.

Finally, we compare the predicted EW(\oiiidoub+\hb), based on our SED fitting, with EW derived empirically at lower and higher redshift: at $z\sim3.8-5.0$ and $z\sim5.1-5.4$, \ha+\niilam\ can be constrained by the IRAC1-IRAC2 color \citep[e.g;][]{shim+11} and \oiiidoub+\hb\ can be constrained with the same color at $z\sim6.6-7.0$ \citep[e.g.,][]{smit+14,castellano+17}. We note again that such empirical constraint at $z\sim6$ is not possible because {\it both} IRAC1 and IRAC2 bands are contaminated by \oiiidoub+\hb\ and \ha, respectively (Fig.~\ref{fig:fig8}). In Fig.~\ref{fig:fig10}, we show the results for the three different SFHs.
Most of the recent studies claim that EW(\oiii+\hb) (or EW(\ha)) is increasing with increasing redshift \citep[e.g.,][]{smit+14,debarros+14}, which would be related to the increase of the specific star-formation rate (sSFR=SFR/\mstar) with redshift, but also possibly related to a change of the ISM physical conditions \citep{faisst+16}, or a change in the IMF. Our comparison show that the EW(\oiii+\hb) predicted by our SED fitting code is consistent with the observed trend. We discuss the BPASS results in Sec.~\ref{sec:bpass}.

\begin{figure}[htb]
\centering
\includegraphics[width=8.75cm,trim=0.5cm 0cm 1.25cm 1cm,clip=true]{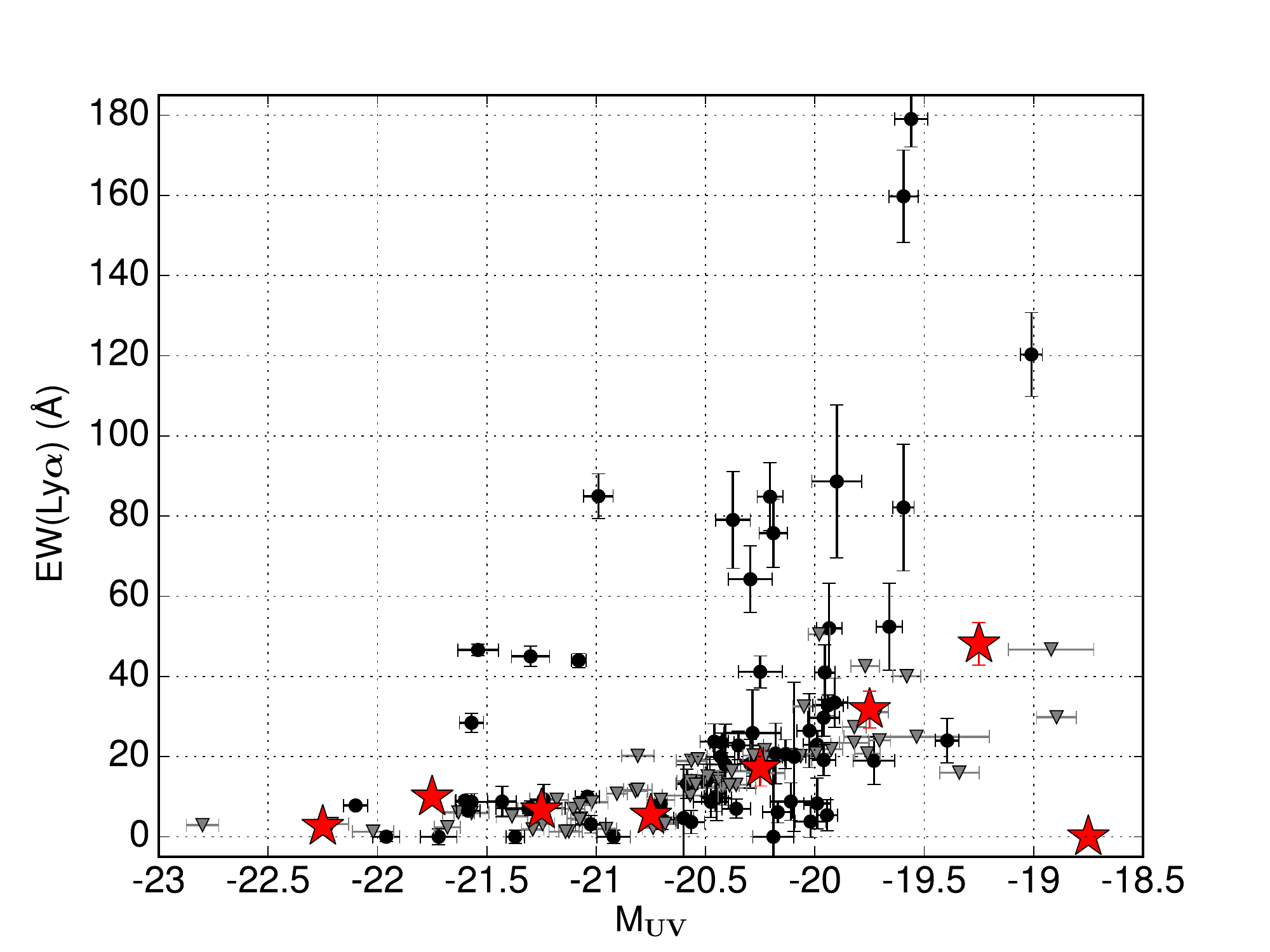}
\caption{EW(\lya) vs. M$_\mathrm{UV}$. We show individual \lya\ detections with black dots and 3$\sigma$ upper limits with grey downward triangles. We also show the average EW in magnitude bins (red stars). Those average values are derived by setting EW(\lya) to zero for all undetected objects that should provide a good approximation of the true relation between EW(\lya) and $\mathrm{M}_\mathrm{UV}$ \citep{schenker+14}.}
\label{fig:fig11}
\end{figure}

\begin{figure}[htb]
\centering
\includegraphics[width=8.75cm,trim=0.5cm 0cm 1.25cm 1cm,clip=true]{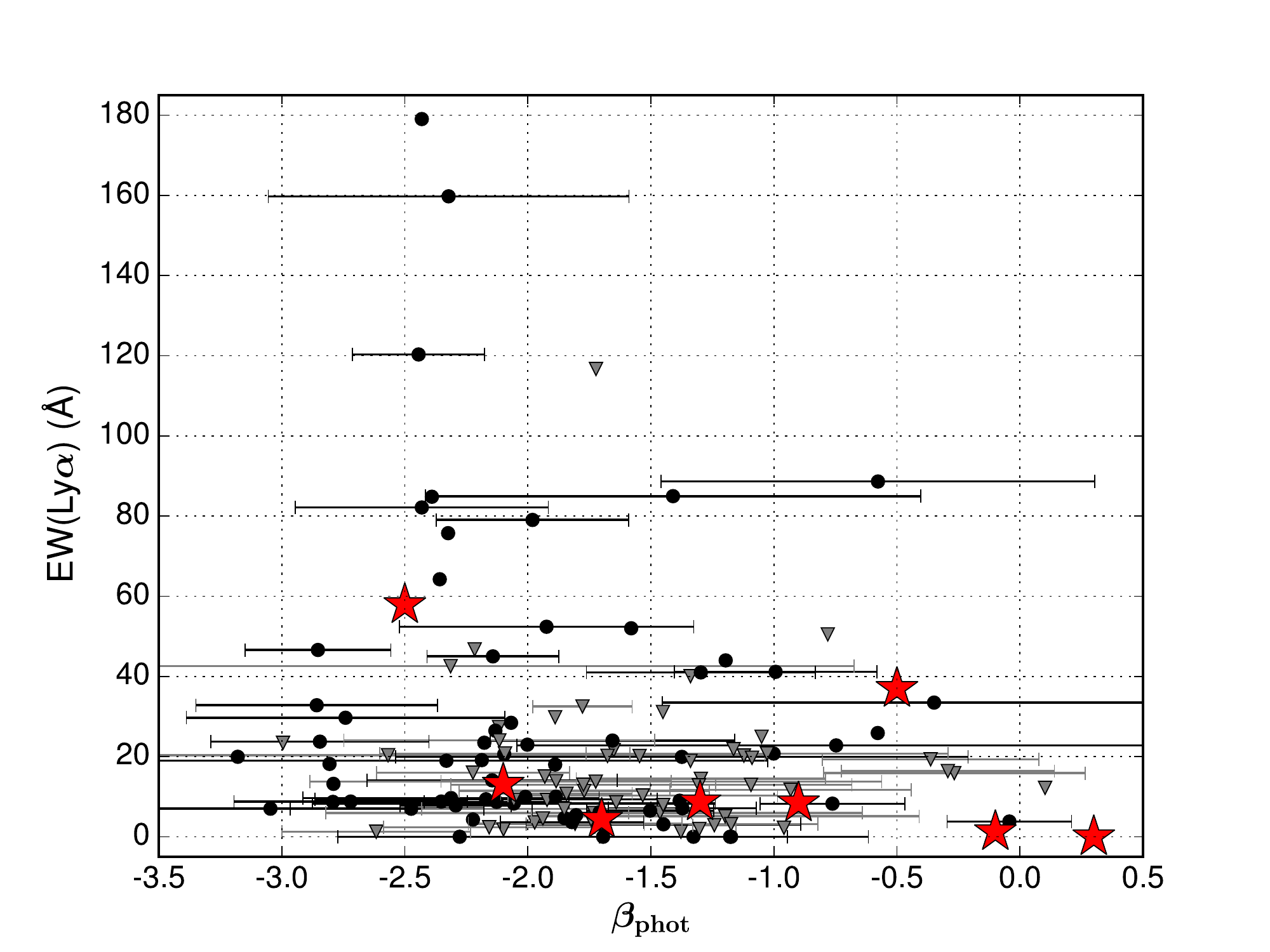}
\caption{EW(\lya) vs. UV $\beta$ slope. Same symbols as in Fig.~\ref{fig:fig11}. To increase the figure clarity, we show $\beta$ errorbars randomly for half of the sample.}
\label{fig:fig12}
\end{figure}

We conclude that while little is known about the nature of the stellar populations (e.g., IMF, SFH, metallicity, binaries/rotation contribution) or the ISM physical conditions at $z\sim6$, our SED fitting procedure is able to provide results consistent with previously observed trends (EW(\oiii+\hb) vs. redshift) and consistent with available SFR tracers.

\begin{figure*}[htb]
\centering
\includegraphics[width=9cm,trim=1cm 1cm 1cm 1cm,clip=true]{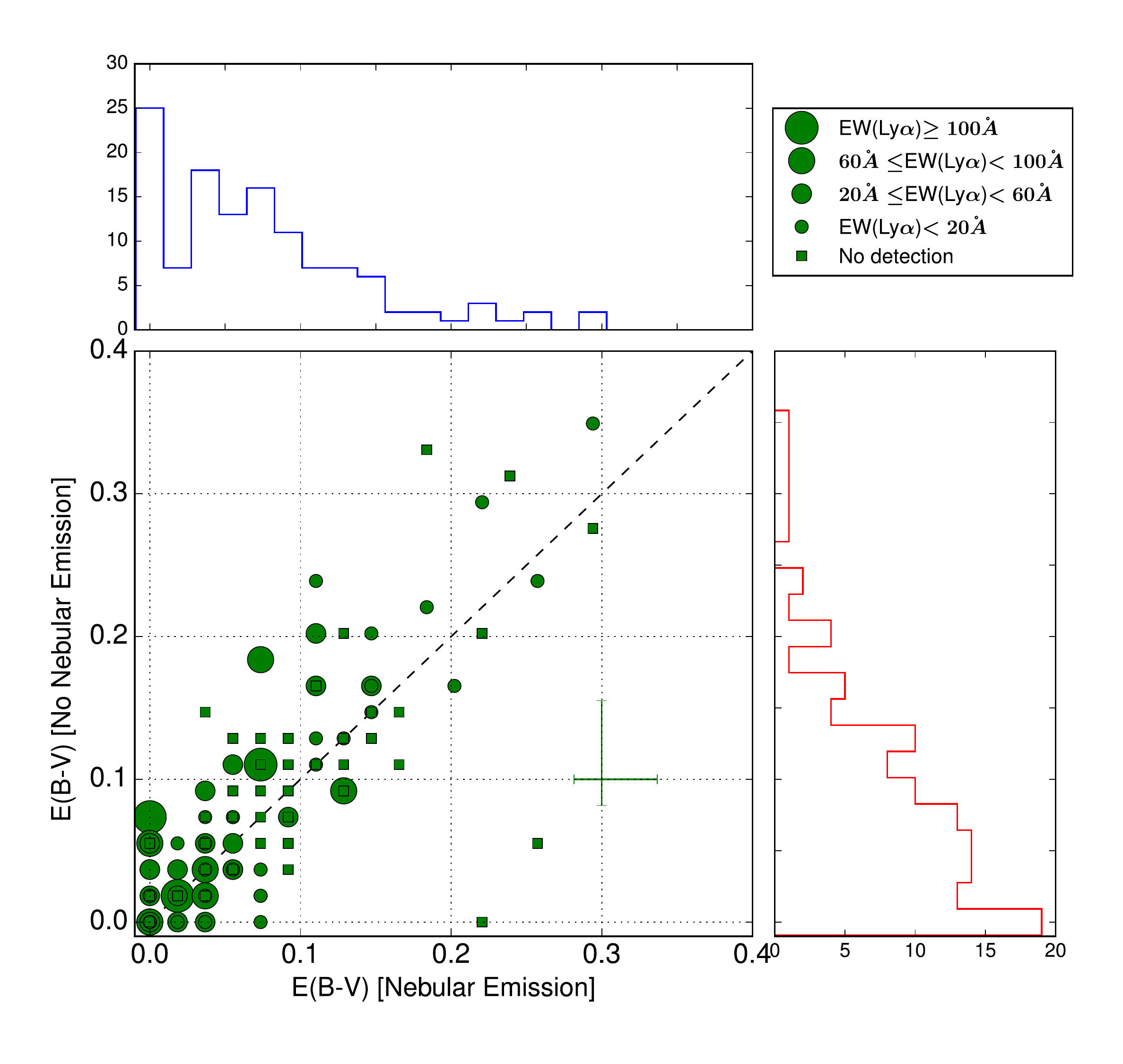}
\includegraphics[width=9cm,trim=1cm 1cm 1cm 1cm,clip=true]{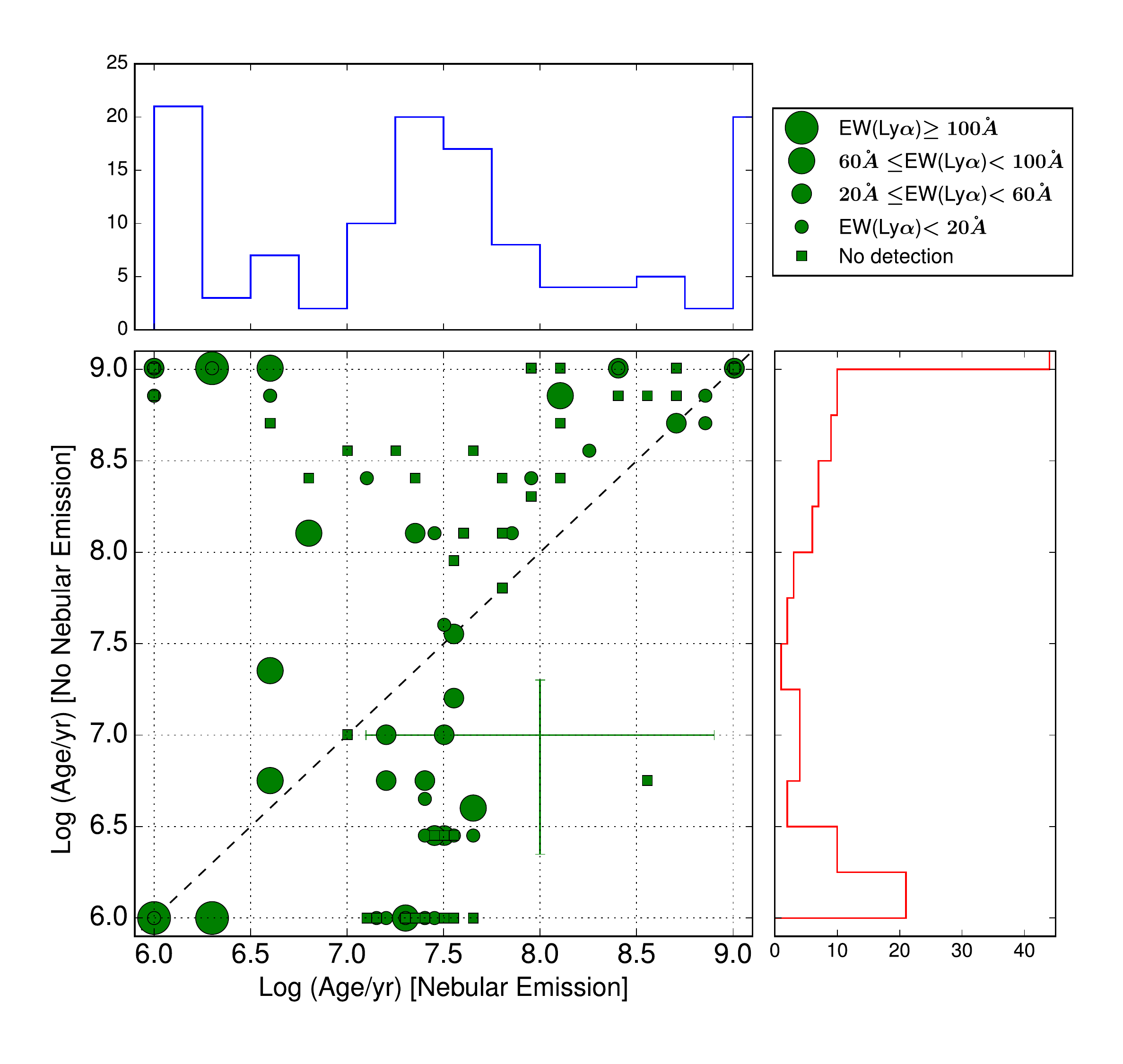}
\includegraphics[width=9cm,trim=1cm 1cm 1cm 1cm,clip=true]{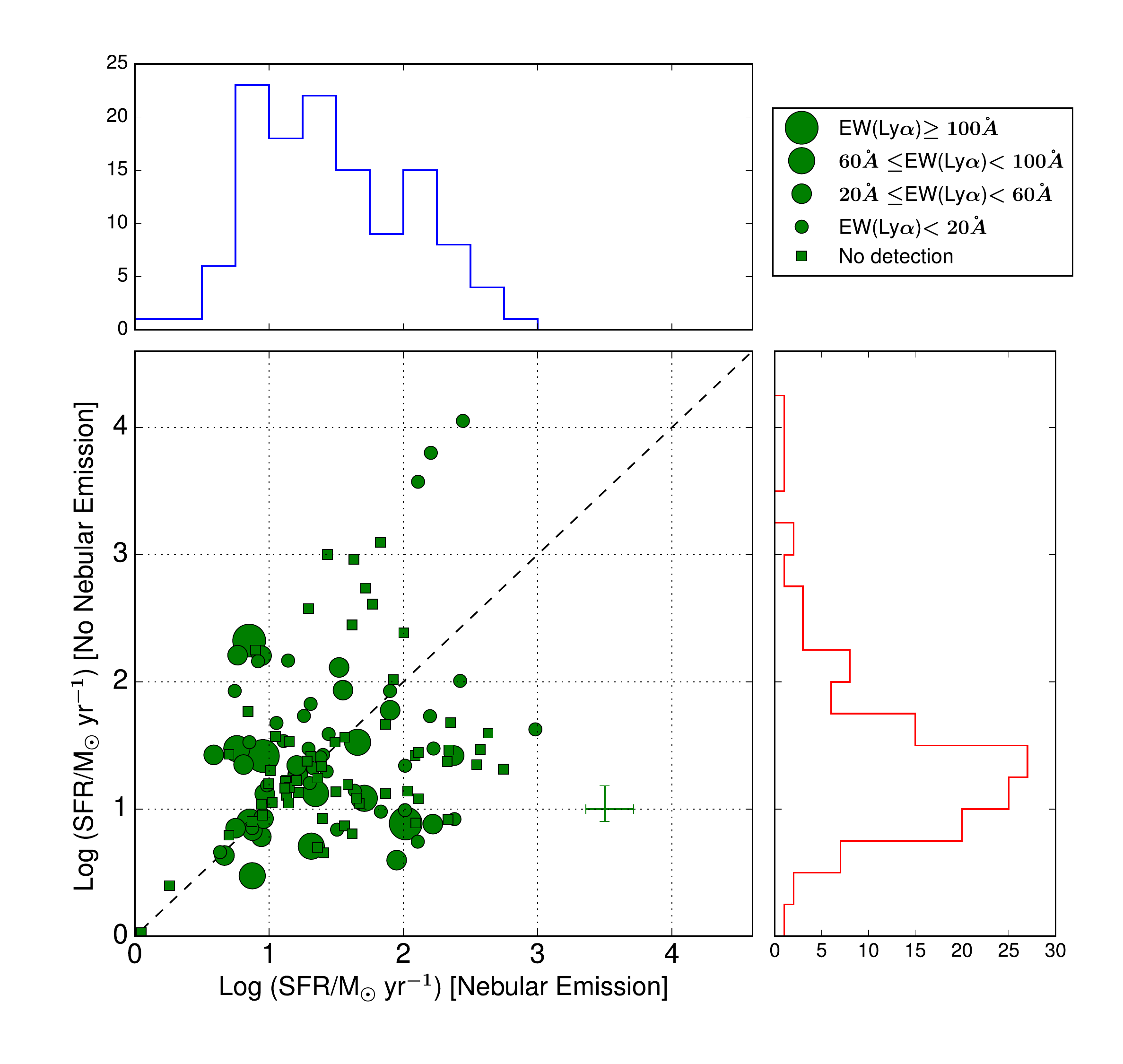}
\includegraphics[width=9cm,trim=1cm 1cm 1cm 1cm,clip=true]{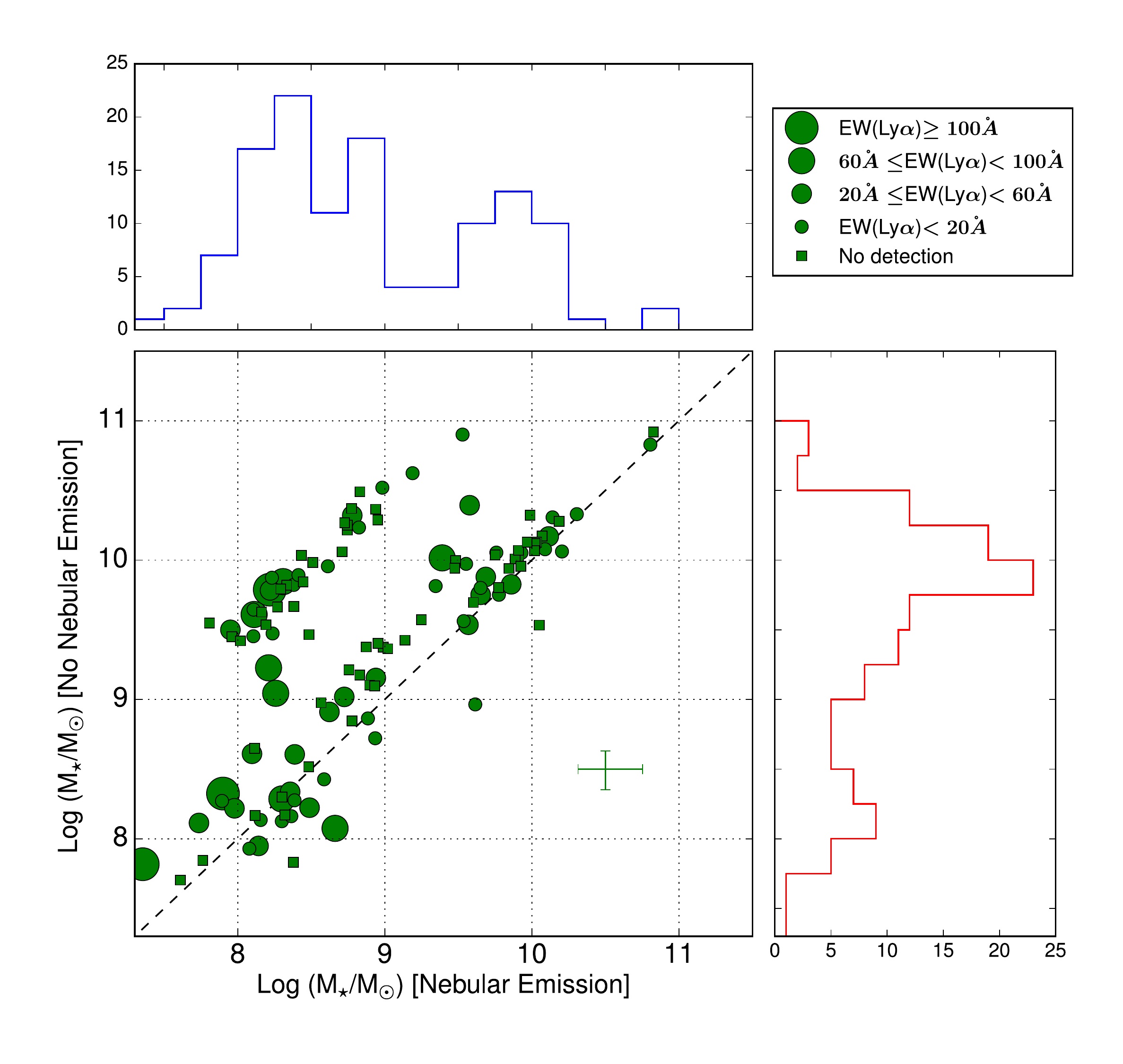}
\caption{Comparison between physical parameters (color excess, age, SFR, stellar mass) derived without accounting and accounting for nebular emission, assuming a constant SFH and a SMC curve. Dashed lines show the one to one relations. The distribution for each parameter is shown on the top of each figure along the x-axis for parameters derived with nebular emission (blue histograms) and on the right side along the y-axis for parameters derived without nebular emission (red histograms). Typical errorbars for each parameter are shown on the lower right side.}
\label{fig:fig13}
\end{figure*}


\subsection{Relation between \lya\ and UV properties}
\label{sec:lyauv}

Several studies either in the local Universe \citep[e.g.,][]{hayes+14} or at high-redshift \citep[e.g.,][]{shapley+03,erb+06a,reddy+06b,pentericci+07,pentericci+10,kornei+10,hathi+16,matthee+16,trainor+16} have found physical differences between Lyman-$\alpha$ emitters and non-Lyman-$\alpha$ emitters \citep[but see also][]{,hagen+16}. The general trend is that UV selected LAEs with the largest EW(\lya) have bluer UV $\beta$ slopes ($f_\lambda\propto\lambda^\beta$), fainter absolute UV magnitude, younger stellar populations, lower stellar masses, lower SFRs, and are less dusty than galaxies with lower EW(\lya). 

We first compare EW(\lya) with two quantities which are not dependent on assumptions: the UV absolute magnitudes and the UV $\beta$ slopes. The absolute UV magnitude $\mathrm{M}_\mathrm{UV}$ refers to the absolute magnitude at 1500\AA\ that we derive by using the integrated SED flux in an artificial filter of 200\AA\ width centered on 1500\AA. We find that UV bright galaxies with large \lya\ equivalent widths are absent, while fainter galaxies exhibit a large EW(\lya) range (Fig~\ref{fig:fig11}). This trend is consistent with results from numerous previous high-redshift studies \citep{ando+06,pentericci+09,schaerer+11,stark+11,cassata+15} and it has been interpreted as the result of the spatial extension of \lya\ emission, since the spatial extension scales with  galaxy size \citep{wisotzki+16}. As the UV $\beta$ slope is an observed property and a proxy for the dust attenuation \citep[e.g.,][]{bouwens+14}, we show in Fig.~\ref{fig:fig12} the relation between EW(\lya) and the UV $\beta$ slope for our sample. $\beta$ slopes are derived directly from the photometry \citep{castellano+12}. The highest EW(\lya) are found for the bluest $\beta$ slopes, while EW(\lya) as high as $\sim80\AA$ can be found for any observed slope value.

The trend between EW(\lya) and UV magnitude is stronger than between EW(\lya) and $\beta$, but the relatively large uncertainties affecting the UV slope derivation (Fig.~\ref{fig:fig12}) prevent from providing a conclusion about a possible stronger dependence of EW(\lya) on the UV magnitude. Nevertheless, compared to previous samples \citep[e.g.,][]{stark+11}, we find similar relations between \lya\ and UV properties.

\subsection{Relation between \lya\ and other physical properties}
\label{sec:lyaphys}

Thanks to the nebular emission modeling we can add an additional constraint to our SED fitting procedure: for each galaxy, we compare predicted \lya\ fluxes from SED fitting with the observed ones, and we exclude solutions predicting ``intrinsic" \lya\ fluxes (i.e., SED predicted) lower than observed fluxes.
Using this method, \sfrlya\ and \sfrsed\ are consistent for any SFH, including declining ones.

We show the relations between EW(\lya) and physical properties (age, stellar mass, color excess, SFR) in Figure~\ref{fig:fig13}. In all cases, we assume a constant SFH and a SMC attenuation curve (we find qualitatively the same results for any set of assumptions described in Sec.~\ref{sec:sed}).

Accounting for nebular emission at $z\sim6$ leads to best fit parameters with lower stellar masses and younger ages \citep{schaererdebarros09}, but when accounting for uncertainties, the result is less clear regarding ages with typical error of $\pm1\mathrm{dex}$. Age estimation depends on the assumed SFH and on the fit of both the UV $\beta$ slope and a color constraining the Balmer break, for example, at $z\sim6$, the $K-\mathrm{IRAC1}$ color. Using our models with nebular emission (assuming zero dust attenuation), $K-\mathrm{IRAC1}$ varies between 0.40 and 0.90, and it goes from -0.85 to 0.70 with no nebular emission.  Another difference between the two models is that $K-\mathrm{IRAC1}$ increases from 1Myr to 1Gyr in models without nebular emission (because of the increasing Balmer break), while it decreases from 1Myr to $\sim50$Myr ($0.39<K-\mathrm{IRAC1}<0.84$) and increases from $\sim50$Myr to 1Gyr ($0.39<K-\mathrm{IRAC1}<0.77$) when accounting for nebular emission. This is explained by the fact that from 1Myr to 50Myr, the $K-\mathrm{IRAC1}$ color is dominated by strong emission lines affecting the IRAC1 channel (\oiii+\hb), and between 50Myr to 1Gyr, there is an increasing impact of the Balmer break. The median $K-\mathrm{IRAC1}$ color of our sample ($0.71^{+1.01}_{-0.49}$) can be explained by relatively old ages for models without nebular emission or by a large range of ages for models accounting for nebular emission. To attempt to break this degeneracy between age and emission line equivalent width, we define the effective escape fraction \fefflya\ as the ratio between the observed \lya\ flux to the SED predicted \lya\ flux and this quantity is the result of the combined effect from \lya\ radiative transfer in the ISM, and the effect of the CGM and intergalactic medium \citep{nagamine+10,dijkstrajeeson13}. Assuming $\fefflya\leq1$ in the SED fitting procedure, we break the degeneracy between age and EW(\oiii+\hb) for galaxies with EW(\lya)$>80\AA$, and the number of acceptable fit is reduced for galaxies with $40\AA<\mathrm{EW}(\lya)<80\AA$ while not affecting significantly the best-fit parameters. Using this additional constraint ($\fefflya\leq1$), we find that the relation between EW(\lya) and age is similar to the one observed at low-z with the galaxy exhibiting largest EW(\lya) being the youngest  \citep[e.g.,][]{hayes+14}. Regarding other physical parameters, we find that the LAEs with the largest EW(\lya) are less dusty, less massive, and less star-forming than non-LAEs or LAEs with lower EW(\lya). The relations between EW(\lya) and physical parameters are similar at low and high-redshift and suggests that the escape of \lya\ photons is likely driven by similar physical processes.

We find that $z\sim6$ galaxies in our sample have a median stellar mass $\log(\mstar/\msun)=8.7^{+1.3}_{-0.7}$, an age of $25^{+769}_{-24}\mathrm{Myr}$, an instantaneous star-formation rate $\log(\mathrm{SFR}/\msunyr)=1.4^{+0.9}_{-0.7}$, and a color excess $\mathrm{E(B-V)}=0.06^{+0.11}_{-0.06}$ (assuming an SMC curve, twice this value for a Calzetti or Reddy curve). If we define a LAE as a galaxy with $\mathrm{EW}(\lya)\geq20\AA$, then we find than non-LAE have typical properties similar to the average of the sample (they made up 75\% of the sample), while LAEs are slightly less massive, less star-forming, and have a higher specific SFR ($\mathrm{SFR}/\mstar$). The main difference between LAEs and non-LAEs is the dust extinction, with LAEs having color excesses twice as small as the typical value for non-LAEs. Again, this relation between dust extinction and the ability for \lya\ photons to escape have been found in previous studies at lower redshift \citep[e.g.,][]{pentericci+07,verhamme+08,atek+09,hayes+11}.

We also derive the median \fefflya\ for our sample with $\fefflya=0.23^{+0.36}_{-0.17}$. This result is mostly independent from the model assumed. Our $\fefflya$ is remarkably consistent with values derived at by comparing the \lya\ and UV luminosity functions $z\sim6$ \citep{hayes+11,blanc+11,dijkstrajeeson13}.
We discuss the implication of this result in Sec.~\ref{sec:lyaigm}.

\subsection{Effect of spectral synthesis models accounting for binary stars}
\label{sec:bpass}

Until recently the effects of binary stars and stellar rotation were neglected in stellar population synthesis models \citep[e.g.,][]{elridge+08,elridgestanway09,levesque+12}. Models taking into account these effects are able to fit young local star-clusters \citep{wofford+16} and are necessary to reproduce nebular emission lines not typically observed in local galaxies (e.g., \civ, \ciii, and \heii) but that seem more common at high-z \citep[e.g.,][]{shapley+03,stark+14b,steidel+16,vanzella+16b,amorin+17,smit+17,vanzella+17}. These lines require harder ionizing spectra and cannot be reproduced by models not taking into account binaries or rotation. Indeed, the main difference between standard BC03 templates used to perform SED fitting in this work and models accounting for binaries and/or rotation is an increased ionizing flux and a harder ionizing flux \citep{stanway+16}. Those models with an increased ionizing photon output have also been favored recently because of the current stringent constraints on the typical Lyman continuum escape fraction of star-forming galaxies \citep[e.g.,][]{grazian+16} that are difficult to reconcile with a realistic scenario for the cosmic reionization where star-forming galaxies are thought to be the main contributors to the ionizing background \citep{bouwens+16a}.

The Binary Population and Spectral Synthesis code \citep[BPASS;][]{elridgestanway09} was developed specifically to take into account binary evolution in modelling the stellar populations. We use the BPASSv2\footnote{http://bpass.auckland.ac.nz/2.html} models with $Z=0.2Z\sun$ for a constant SFH to fit our sample and we compare the results with those obtained with BC03 templates. An example of a fit is shown in Fig.~\ref{fig:fig8}. BPASS models are able to reproduce UV slopes as well as BC03 (Sec.~\ref{sec:sed}) and the physical parameters are similar to those derived with BC03, except for the stellar mass. As shown in Figure~\ref{fig:fig10}, because BPASS models have a larger ionizing photon output, emission line fluxes and equivalent widths at a given age and dust extinction are larger, relative to BC03 results. EW(\oiii+\hb) are more than twice as large with BPASS than with BC03. While the trend of EW(\oiii+\hb) with redshift is uncertain, mainly because of the gap between $z\sim5.5$ and $z=8.0$ with no constraints on equivalent width except for the small samples from \cite{smit+14} and \cite{castellano+17}, it seems that the values that we derive for our $z\sim6$ sample using either BC03 or BPASS templates are consistent with expectations regarding results at $z<6$ and result at $z\sim8$ from \cite{labbe+13}.
While results obtained with BPASS and BC03 are consistent within their uncertainties, the large EWs found with BPASS have an impact on stellar mass estimation: on average, stellar masses are 0.4dex lower using BPASS templates\footnote{We compute the mass normalization of BPASSv2 assuming a 30\% mass fraction recycled in the ISM as in \cite{castellano+17}.}. Another effect of using BPASS templates is that the predicted \lya\ fluxes are larger and the effective escape fraction is in this case $\fefflya=0.15^{+0.24}_{-0.10}$. This lower \fefflya\ value seem in greater tension with values from literature than the \fefflya\ derived in Sec.~\ref{sec:lyaphys}, but it is still consistent within uncertainties with the values from other studies \citep{dijkstrajeeson13}.

Using BPASS templates, our results remain globally unchanged except regarding the emission lines strength, because of the increased ionizing output. Accordingly the stellar masses are decreased because of the lines contribution to IRAC1 and IRAC2. 
We cannot conclude about the accuracy of the binary modeling in BPASS and this will have to be tested with {\it JWST} observations \citep{stanway17}. A possible test would be to perform SED fitting of high-z galaxies with BPASS at a redshift where empirical constraints on emission line EWs are available \citep[e.g., $z\sim4$;][]{shim+11}.

\section{Discussion}

\subsection{IGM effect on \lya\ visibility at $z\sim6$}
\label{sec:lyaigm}

Several studies have tried to derive the impact of the IGM on the \lya\ visibility in a fully ionized Universe. In \cite{dijkstra+07}, \cite{zheng+10}, and \cite{laursen+11}, the IGM transmission to \lya\ (\tigm) is found to be low with $\tigm\leq0.01-0.3$ at $z\sim6$. Values as low as $\tigm=0.01$ cannot be reconciled with the effective escape fraction found in our work $\fefflya=0.23^{+0.36}_{-0.17}$, but higher values ($\tigm\sim0.3$) would be consistent. We have defined the {\it effective} \lya\ escape fraction as the result of the combined effects of ISM, CGM and IGM, and the {\it relative} \lya\ escape fraction is defined as the result of ISM only, then
\begin{equation}
\fefflya=\tigm\times\flyarel
\label{eq:tigm}
\end{equation}
Therefore to constrain \tigm, we need to have constraints on \flyarel. Several relations have been derived between E(B-V) and the effective \lya\ escape fraction from $z\sim0$ to $z\sim3$ \citep[e.g.,][]{verhamme+08,atek+09,kornei+10,hayes+11,yang+17} and while the definition of E(B-V) varies among studies, they all find that \fefflya\ decreases with increasing color excess. Interestingly, at those lower redshifts, the effect of the IGM on the \lya\ visibility is expected to be much lower \citep[$\tigm\sim0.8$ at $z\sim3.5$;][]{laursen+11} and so $\fefflya\sim\flyarel$. While the large uncertainties on our derived physical parameters like E(B-V) and \fefflya\ for individual galaxies preclude any attempt to derive similar relations with our data, we stress that {\it on average} we find that the main difference between LAEs and non-LAEs is the dust extinction (Sec.~\ref{sec:lyaphys}). This result suggests that the processes governing \lya\ escape from galaxies at low redshift are similar to those at high redshift. Thus to put constraints on \tigm, we assume that there is a relation between E(B-V) and \flya\ at $z\sim6$ similar to those found at low redshift.

The first difficulty arises from the choice of the relation that we want to use at $z\sim6$. For example, in \cite{atek+09}, E(B-V) is derived assuming the \cite{cardelli+89} attenuation curve and using the Balmer decrement, while \cite{hayes+11} use stellar color excesses derived from SED fitting and a \cite{calzetti+00} curve, finally \cite{yang+17} add a term based on \lya\ velocity red-peak on the relation between E(B-V) and \fefflya. 

We choose to use the relation described by Eq.~4 in \cite{hayes+11} because the E(B-V) values are derived in the same way as in the present work. This relation is:
\begin{equation}
\flyarel=C_\mathrm{Ly\alpha}\times10^{-0.4\times\mathrm{E(B-V)}\times k_\mathrm{Ly\alpha}}
\label{eq:hayes}
\end{equation}
\cite{hayes+11} derive a value of $C_\mathrm{Ly\alpha}=0.445$ and $k_\mathrm{Ly\alpha}=12$ using a Calzetti curve.

However, due to numerous uncertainties about $z\sim6$ galaxies like the ionizing output of the stellar population or the dust attenuation curve \cite[][]{bouwens+16b}, we test different sets of assumptions to derive the average color excess E(B-V) of our sample, using both BC03 and BPASS templates with Calzetti and SMC curves. The difference in terms of average color excess between BPASS and BC03 is negligible, but the choice of attenuation curve introduces a factor 2 difference, with the Calzetti curve leading to larger color excess. Using Eq.~\ref{eq:hayes}, we get $\flyarel=0.08^{+0.12}_{-0.03}$ (BC03+Calzetti), $\flyarel=0.06^{+0.18}_{-0.03}$ (BPASS+Calzetti), and $\flyarel=0.22^{+0.06}_{-0.15}$ (BC03+SMC and BPASS+SMC). Then we use Eq.~\ref{eq:tigm} with the two values derived for \fefflya\ in Sec.~\ref{sec:lyaphys} and \ref{sec:bpass},  using BC03 and BPASS templates respectively, to put constraints on the IGM transmission to \lya\ photons. Accounting for all the models considered, we obtain $\tigm\geq0.18$. 
This value is relatively high in comparison with most of the theoretical studies that derived \tigm, however accounting for outflows, \cite{dijkstra+11} found such high \tigm\ at $z\sim6$ ($\tigm>0.50$). Therefore the lower limit that we find for the IGM transmission would be easily explained if outflows are ubiquitous in $z\sim6$ galaxies, which seems consistent with current high-redshift observations \citep{stark+17}, while possibly with a lower velocity than at $z\sim2$ \citep{pentericci+16}. \cite{dijkstra+11} stress than even relatively low outflow velocities can be sufficient to have a large IGM transparency to \lya\ photons.


 \section{Conclusions}
 \label{sec:conclu}

In this work, we report deep observations of an $i$-dropout sample with VLT/FORS2 to search for \lya\ emission. The dropout selection has been designed to avoid bias toward faint LAEs with large EW(\lya). Combining our data with archival data, we construct a large star-forming galaxy sample spectroscopically confirmed at $z\sim6$, with 127 galaxies with redshift confirmed either by detecting \lya\ emission/continuum emission (Lyman break) or by excluding low redshift solution through the lack of detection of \ha, \oiiidoub, or \oiilam. All our galaxies are $H$-band detected. Thanks to the size of our sample covering five fields, we determine a new $z\sim6$ LAE fraction, minimizing cosmic variance.
We derive physical properties using SED fitting while we minimize the number of assumptions going into our analysis to compare the properties of LAEs and non-LAEs, derive the effective escape fraction \fefflya, and constrain the IGM transmission to \lya\ photons \tigm.

In summary, we find:

\begin{enumerate}
\item The median LAE fractions for bright and faint galaxies in our sample are lower  than found in previous  studies \cite[e.g.,][]{stark+11}, while still consistent with results reported in the literature within uncertainties.
\item  Our data are consistent with a drop or a flattening of the relation between the LAE fraction and redshift at $5<z<6$. This can be a sign of an already increasing IGM neutral fraction at $z<6$.
\item By comparing \sfrsed\ with \sfrlya, we find that declining star-formation histories underestimate the star-formation rate for $~50\%$ of our sample spectroscopically detected.
\item Our sample exhibits the same trends between EW(\lya) and $\mathrm{M}_\mathrm{UV}$, and between EW(\lya) and the UV $\beta$ slopes as previously reported at lower redshift \citep[e.g.,][]{pentericci+09}: the largest EW(\lya) will be found for the faintest and bluest galaxies. LAEs are slightly less massive and less star-forming than non-LAEs but those differences are well within the uncertainties. The main difference is the dust extinction with an average color excess for non-LAEs twice as large as the average LAE color excess. These results are mostly independent from assumptions.
\item We test stellar templates incorporating the effect of binaries (BPASSv2) and they lead to similar physical properties except for increased nebular emission fluxes due to a higher ionizing photon output, and accordingly an increase of EW(\oiii+\hb) which leads to an average decrease of the stellar mass by $\sim0.4\mathrm{dex}$.
\item By comparing observed \lya\ luminosities with SED predicted \lya\ luminosities, we derive an effective escape fraction of $\fefflya=0.23^{+0.36}_{-0.17}$, consistent with values derived by comparing UV and observed \lya\ luminosities \citep{blanc+11,hayes+11}.
\item Assuming that physical processes governing the escape of \lya\ photons from galaxies are similar at low- and high-redshift, we derive a lower limit to the IGM transmission to \lya\ photons $\tigm\gtrsim0.20$. Such large IGM transmission is expected if outflows are present \citep{dijkstra+11} which is also consistent with current constraints \citep{pentericci+16,stark+17}.
\end{enumerate}

\begin{acknowledgements}
This paper has been greatly improved by the referee comments and suggestions.
We thank Anne Verhamme, Pascal Oesch, Mark Dijkstra, Matthew Hayes, and Daniel Schaerer for useful discussions and suggestions. RM acknowledges support by the Science and Technology Facilities Council (STFC) and the ERC Advanced Grant 695671. QUENCH.
\end{acknowledgements}

\bibliographystyle{aa}
\bibliography{ref}

\end{document}